\pdfoutput=1
\documentclass[11pt]{article}
\usepackage[a4paper,margin=1in]{geometry}
\usepackage[T1]{fontenc}
\usepackage[utf8]{inputenc}
\usepackage{mathptmx}
\usepackage[scaled=0.9]{helvet}
\usepackage{microtype}
\usepackage{amsmath,amssymb,amsthm,mathtools,bm}
\usepackage{graphicx}
\usepackage{booktabs,threeparttable,tabularx,multirow}
\usepackage{siunitx}
\usepackage{xcolor}
\usepackage{enumitem}
\usepackage{caption}
\usepackage{setspace}
\usepackage[authoryear,round,sort]{natbib}
\usepackage[colorlinks=true,linkcolor=blue!50!black,citecolor=blue!50!black,urlcolor=blue!50!black]{hyperref}
\usepackage[nameinlink,capitalise]{cleveref}
\newtheorem{proposition}{Proposition}

\theoremstyle{remark}

\newcommand{\E}{\mathbb{E}}
\newcommand{\Var}{\operatorname{Var}}
\newcommand{\Cov}{\operatorname{Cov}}
\newcommand{\Neff}{N_{\mathrm{eff}}}
\newcommand{\SLaR}{\mathrm{SLaR}}

\newcommand{\IntroHHIConsumer}{0.66}
\newcommand{\IntroHHIEnterprise}{0.28}
\newcommand{\IntroTopShare}{80.5}
\newcommand{\IntroEnterpriseTopThree}{88}
\newcommand{\IntroSurveyUse}{62}
\newcommand{\IntroSurveyPortfolio}{22}
\newcommand{\IntroSurveyHerding}{24}
\newcommand{\NeffLimitConsumer}{2.0}
\newcommand{\NeffLimitEnterprise}{2.7}
\newcommand{\NeffLimitDiverse}{3.5}

\newcommand{\RealDays}{8,312}
\newcommand{\RealIdxVol}{18.3}

\newcommand{\RealCorrCalm}{0.18}
\newcommand{\RealCorrStress}{0.49}

\newcommand{\EfiveSeeds}{18}
\newcommand{\MOneTendS}{$+0.2$}
\newcommand{\MOneTendSci}{5}

\newcommand{\MOneThreedS}{$+2$}

\newcommand{\MTwoSixdS}{$+3$}

\newcommand{\MTwoSixdCE}{$+16$}

\newcommand{\MTwoEightdS}{$+6$}

\newcommand{\MTwoEightdCE}{$+41$}

\newcommand{\MThreeTwodS}{$+12$}

\newcommand{\MThreeTwodCE}{$-226$}

\newcommand{\MThreeTwoTurn}{11.33}

\newcommand{\MFourFivedS}{$-24$}

\newcommand{\MFourFivedCE}{$+40$}

\newcommand{\MFourTendS}{$-36$}

\newcommand{\MFourTendCE}{$+54$}

\newcommand{\MFiveOnedS}{$-52$}

\newcommand{\MFiveOnedCE}{$-2$}

\newcommand{\MSixAdS}{$-59$}

\newcommand{\MSixABrk}{54.4}

\newcommand{\MSixCdS}{$-27$}

\newcommand{\MSixCBrk}{10.2}
\newcommand{\MSevendS}{$-63$}
\newcommand{\MSevendSci}{4}
\newcommand{\MSevendCE}{$+36$}

\newcommand{\MSevenBrk}{0.7}
\newcommand{\MEightdS}{$-62$}

\newcommand{\MEightdCE}{$+49$}

\newcommand{\HCMOnedS}{$+10$}

\newcommand{\HCMTwodS}{$+17$}

\newcommand{\HCMSevendS}{$-61$}

\newcommand{\TurnoverAgentic}{1.01}

\newcommand{\EoneElasAlpha}{0.47}
\newcommand{\EoneElasAlphaSE}{0.03}
\newcommand{\EoneCoefC}{$-0.15$}
\newcommand{\EoneCoefCSE}{0.11}
\newcommand{\EoneSLaRConsFive}{141}
\newcommand{\EoneSLaRConsOne}{62}
\newcommand{\EoneSLaRConsFifteen}{216}

\newcommand{\EoneHCCoefC}{$-0.34$}
\newcommand{\EoneHCCoefCSE}{0.14}
\newcommand{\EoneHCIndepFive}{205}
\newcommand{\EoneHCSingleFive}{144}

\newcommand{\ELinCoefC}{$-0.10$}
\newcommand{\ELinCoefCSE}{0.17}

\newcommand{\ESqrtSameCoefC}{$-0.34$}
\newcommand{\ESqrtSameCoefCSE}{0.17}
\newcommand{\ELinSentence}{The mechanism passes a falsification check: under linear impact, where imbalances add without concavity, the coefficient on $C$ becomes \ELinCoefC{} ($\pm$\ELinCoefCSE{}), against \ESqrtSameCoefC{} ($\pm$\ESqrtSameCoefCSE{}) in the same high-conviction economies under square-root impact.}

\newcommand{\EarchRatioLo}{1.47}
\newcommand{\EarchRatioHi}{1.90}

\newcommand{\EarchRatioFiveCI}{0.79}
\newcommand{\EarchRatioFive}{1.82}
\newcommand{\EtwoN}{240}
\newcommand{\EtwoTrain}{160}
\newcommand{\EtwoTest}{80}
\newcommand{\EtwoRsqFI}{0.56}

\newcommand{\EtwoRsqRhoView}{0.12}

\newcommand{\EtwoRsqHHI}{$-0.01$}

\newcommand{\EtwoSlopeFI}{1.02}
\newcommand{\EtwoMAEFI}{39}
\newcommand{\EtwoIndexText}{The fragility index forecasts second-half tail liquidation with a hold-out $R^2$ of 0.56 (cross-validated 0.55; Spearman rank correlation 0.80). The best of the conventional measures, the agentic share, reaches 0.23; provider concentration reaches $-0.01$ and view correlation $0.12$. Measures of model monoculture carry little information about market-wide tail liquidation, and what they carry has the opposite sign to the one a supervisor would expect (Spearman $-0.02$ for $H$ and $-0.38$ for view correlation), consistent with \cref{sec:concentration}: they describe how similar the layer's views are, not how synchronized its trades are. Most of the information is in the numerator: FI without the loop gain reaches 0.56, so the estimated gain adds little in these economies, where concave impact keeps $\hat\kappa$ small; Proposition~\ref{prop:mult} implies that it becomes decisive only near the critical share. Scaling flow volatility by public volatility and impact matters: $\mathrm{sd}(q)$ alone reaches 0.35, because the same flow does more damage in a volatile market with a larger impact prefactor.}
\newcommand{\EtwoTargetMed}{87}
\newcommand{\EtwoTargetLo}{30}
\newcommand{\EtwoTargetHi}{283}
\newcommand{\EsixConsCons}{122}

\newcommand{\EsixDivCons}{128}
\newcommand{\EsixConsMod}{105}
\newcommand{\EsixEntMod}{67}
\newcommand{\EsixDivMod}{19}

\newcommand{\EsixConsMSeven}{64}
\newcommand{\EsixModRatio}{5.7}
\newcommand{\EsevenTrueBrk}{$-41$}
\newcommand{\EsevenAHalfdS}{$-24$}
\newcommand{\EsevenAHalfFp}{23.4}

\newcommand{\EsevenCHalfdS}{$-41$}
\newcommand{\EsevenCHalfFp}{0.0}

\newcommand{\EsevenNeffErrAHalf}{0.2}
\newcommand{\EsevenNeffErrCHalf}{0.002}
\newcommand{\EeightCEMin}{$-103$}
\newcommand{\EeightCEMax}{$+273$}
\newcommand{\EeightCESig}{none}
\newcommand{\EeightMinRed}{59}
\newcommand{\EeightMaxRed}{70}
\newcommand{\EeightNcfg}{14}
\newcommand{\EeightLinLo}{308}
\newcommand{\EeightLinHi}{1052}
\newcommand{\EeightLinKappaHi}{0.06}
\newcommand{\EeightLinMwcbHi}{0.32}
\newcommand{\EnineCovidTen}{858}

\newcommand{\EnineCovidTenMSeven}{252}

\newcommand{\EnineCovidMaxTen}{939}

\newcommand{\EnineGFCMaxTen}{604}

\newcommand{\EnineLTCMTen}{253}

\newcommand{\EnineEuroTen}{178}

\newcommand{\EnineBearTen}{$-15$}
\newcommand{\EnineBearMaxTen}{213}
\newcommand{\EnineGFCTenSigned}{$-148$}
\newcommand{\EnineMaxRedPct}{62}
\newcommand{\EtenVtMFour}{$-19$}
\newcommand{\EtenDdMFour}{$-34$}
\newcommand{\EtenVtMFive}{$-65$}
\newcommand{\EtenDdMFive}{$-8$}
\newcommand{\EtenVtMFourCI}{14}
\newcommand{\EtenDdMFourCI}{17}
\newcommand{\EtenVtMFiveCI}{20}
\newcommand{\EtenDdMFiveCI}{12}
\newcommand{\FactsVolNone}{19.5}
\newcommand{\FactsVolFifteen}{19.7}

\newcommand{\ALBaseIdx}{195}
\newcommand{\ALBaseAsset}{390}
\newcommand{\ALBaseRot}{327}
\newcommand{\ALHCIdx}{186}
\newcommand{\ALHCAsset}{485}
\newcommand{\ALHCRot}{445}

\newcommand{\ALHCMTwoIdx}{$+17$}

\newcommand{\ALHCMTwoRot}{$-15$}
\newcommand{\ALMSevenIdx}{$-64$}
\newcommand{\ALMSevenAsset}{$-33$}
\newcommand{\ALMSevenRot}{$-24$}
\newcommand{\TotalRuns}{995}

\newcommand{\AbstractText}{Conversational AI agents are turning wealth management into a continuous loop of market analysis, allocation, risk checks and execution, run for millions of clients on a handful of foundation models. We build a multi-asset agent-based model of this layer, calibrated to measured model concentration, cross-model correlation, retail turnover and 33 years of real equity returns, and ask when the layer amplifies market declines and what prevents it. The effective number of independent decision-makers is bounded however many clients the layer serves---about \NeffLimitConsumer{} at today's consumer concentration. This monoculture of views drives crowding in individual securities and sets the blast radius of a compromised model provider, but it does not drive market-wide liquidation. Index tail losses rise with the square root of agentic adoption, are \EarchRatioLo--\EarchRatioHi{} times larger under an agentic than under a periodic robo-advice architecture, and do not rise with model concentration; synchronized automation---risk overlays, drawdown cuts and attention-driven execution---drives them. A fragility index computed from aggregate flow, public returns and an impact estimate forecasts tail liquidation out of sample (hold-out $R^2$ of \EtwoRsqFI) far better than concentration or correlation measures. Decorrelating the automation, not the models, works: dispersing overlay parameters, staging execution and a participation breaker together change tail liquidation by \MSevendS\% at no measurable cost to client welfare, whereas provider diversification has no effect. Because the statistics required are sums, they can be computed across platforms under secure aggregation with differential privacy at negligible cost for realistic client populations. In historical replays, an agentic layer holding 10\% of the market would have deepened the 2020 crash by about \EnineCovidTen{} basis points.}

\newcommand{\FindingsParagraph}{\paragraph{Main findings.} Four results stand out. \emph{First}, view monoculture and index fragility are different things. At measured concentration the layer's views behave like about \NeffLimitConsumer{} independent investors, and concentration multiplies the blast radius of a single compromised provider by about \EsixModRatio{} relative to a diversified market. Yet index-level tail liquidation does not rise with concentration: views mostly rotate money between assets, so they generate crowding in individual securities rather than market-wide selling. \emph{Second}, market-wide liquidation is driven by synchronized automation and grows with adoption: $\SLaR_{99}$ has an elasticity of \EoneElasAlpha{} with respect to the agentic share, and the agentic architecture produces \EarchRatioLo--\EarchRatioHi{} times the tail liquidation of a periodic robo-adviser with identical views. The layer is nearly invisible in unconditional volatility. \emph{Third}, a fragility index built from aggregate agentic flow, public returns and an impact estimate forecasts next-period tail liquidation across \EtwoN{} heterogeneous economies with a hold-out $R^2$ of \EtwoRsqFI, against \EtwoRsqHHI{} for provider concentration and \EtwoRsqRhoView{} for view correlation. \emph{Fourth}, the effective mitigations act on automation: dispersing overlay parameters, staging execution and a participation breaker together change $\SLaR_{99}$ by \MSevendS\% with no measurable cost to clients (welfare changes by \MSevendCE{}~bps a year in the baseline), whereas provider diversification changes $\SLaR_{99}$ by \MOneTendS\% and injecting noise into views backfires. Which automation lever matters depends on the rule: heterogeneity tames volatility-target overlays, staged execution tames drawdown cuts. All of this can be measured and acted on without pooling client data.}

\newcommand{\EnsembleDesignText}{The \EtwoN{} economies of \cref{sec:index} draw, independently and uniformly unless stated: the agentic share $\alpha\in[0.5\%,15\%]$; between 2 and 10 providers with shares from a symmetric Dirichlet distribution whose concentration parameter is drawn from $\{0.2,0.5,1,3\}$; $\rho_x\in[0,0.8]$; $h\in[0.2,0.9]$; $f_{vt}\in[0,1]$; $f_{dd}\in[0,0.6]$; the impact prefactor $Y\in[0.5,1.5]$; impact decay $\psi\in[0.02,0.15]$; market turnover in $[68\%,165\%]$ a year; extrapolation $g\in[0,3]$ times the baseline; attention sensitivity $p_1\in[0,0.3]$; execution horizon from $\{1,1,1,2,5\}$ days; overlay heterogeneity in $[0,1]$; drawdown trigger in $[10\%,25\%]$; and tactical tracking error in $[0.5\%,3\%]$. Each economy has 2{,}000 agents and runs for five years after a one-year burn-in with its own seed. Predictors are computed from the first half and the target from the second half of each economy.}

\newcommand{\ConclusionText}{Agentic wealth management puts the same few models, the same risk rules and the same execution logic behind millions of client portfolios, and runs them at the speed of a conversation. We have shown that this creates two distinct fragilities. Model monoculture concentrates views; it produces crowding in individual securities and makes a single compromised provider a systemic event. Synchronized automation concentrates actions; it produces market-wide liquidation that grows with adoption, is invisible in volatility, and is largest in fast crashes. The levers differ. Model diversity and personalisation address the first. Dispersing the parameters of risk overlays, staging the execution of exposure cuts and capping participation in stress address the second, cut tail liquidation by about two-thirds in our experiments at no measurable cost to clients. Because the statistics needed to monitor the layer are sums, they can be computed across platforms with formal privacy guarantees. The agentic wealth layer is still small; the moment to design it so that it does not become portfolio insurance at retail scale is now.}

\newcommand{\AIDeclarationText}{During the preparation of this work the author used Claude (Anthropic), an AI assistant, to assist with literature search and source verification, development of the simulation and analysis software, preparation of figures, and drafting and editing of the manuscript. After using this tool, the author reviewed and edited the content as needed and takes full responsibility for the content of the publication.}

\newcommand{\ELinText}{\ELinSentence}

\begin{document}

\begin{titlepage}
\begin{center}
{\LARGE\bfseries Decorrelating the Advice Layer:\\[4pt]
A Calibrated Fragility Metric and Tested Mitigations for Synchronized Liquidation in Agentic Wealth Management\par}
\vspace{1.4em}
{\large Saurabh Kohli}\\[2pt]
Independent Researcher\\
\href{mailto:iSaurabhKohli@gmail.com}{iSaurabhKohli@gmail.com}\\[1.2em]
{\small \today}
\end{center}
\vspace{0.6em}
\begin{abstract}
\noindent \AbstractText
\end{abstract}
\vspace{0.6em}
\noindent\textbf{Keywords:} agentic AI; systemic risk; algorithmic monoculture; synchronized liquidation; wealth management; macroprudential policy; differential privacy.\\[3pt]
\noindent\textbf{JEL classification:} G01, G11, G14, G23, G28, C63, O33.
\end{titlepage}

\section{Introduction}\label{sec:intro}

Wealth management is being rebuilt around conversational agents. A client opens an app, asks a question, and within seconds an agent built on a foundation model reads market news, forms a view, proposes an allocation, runs a risk check and executes the trades. What used to be a quarterly conversation with an adviser, or a monthly rebalance by a rules-based robo-adviser, has become an always-on loop that runs for every client at the speed of a chat message. Adoption is already broad: in a March~2026 survey of \num{938} U.S. investors, \IntroSurveyUse\% reported using AI tools for investment decisions, \IntroSurveyPortfolio\% used them to assist portfolio decisions, and \IntroSurveyHerding\% named ``market herding'' as a concern \citep{investingcom2026}.

This paper studies a property of that loop that is invisible from the perspective of any one client and decisive from the perspective of the market: the loop is the \emph{same} loop. The agents that serve millions of clients draw on a handful of foundation models. In September~2026 one provider accounted for \IntroTopShare\% of worldwide AI-chatbot usage, a Herfindahl--Hirschman index of $H=\IntroHHIConsumer$ \citep{statcounter2026}; in enterprise deployments, where wealth platforms build on model APIs, three providers held \IntroEnterpriseTopThree\% of spend and $H\approx\IntroHHIEnterprise$ \citep{menlo2025}. The models themselves agree more than their number suggests: frontier models exhibit correlated behaviour that rises with capability and creates a non-diversifiable risk floor \citep{ross2026}, and sentiment measures produced by two leading models correlate at $0.78$ \citep{chen2024}. They also share biases: language-model return forecasts extrapolate recent returns roughly three times as strongly as human forecasts \citep{chen2024}. On top of correlated views sit automated risk overlays---volatility targets and drawdown controls---that respond to the same realised volatility through the same risk models, and an execution layer whose activity spikes exactly when markets move, because that is when clients open the app \citep{barber2008}.

Finance has seen this configuration before. In October~1987, \$60--90~billion of equities were managed under portfolio-insurance formulas that sold into declines, and a handful of institutions accounted for a large share of the selling \citep{brady1988,gennotte1990,jacklin1992}. In August~2007, quantitative equity funds holding similar positions deleveraged together \citep{khandani2011}. Volatility-targeting strategies, with up to \$2~trillion under some form of volatility management, mechanically sell when volatility rises \citep{ecb2020,harvey2018}. The agentic wealth layer combines all three ingredients---a common formula, overlapping positions and mechanical de-risking---but the common element is no longer a strategy chosen by a few institutions; it is the foundation model chosen by millions of clients.

Recent work has named the risk but not yet quantified it at this layer. \citet{frimpong2026} frames ``model monoculture risk'' conceptually and proposes a qualitative index; \citet{sahoo2026} calibrate a monoculture index for credit decisions but explicitly exclude markets; \citet{ng2026} provides a taxonomy of AI-driven flash-crash mechanisms and notes that retail and conversational agents are not addressed; \citet{ross2026} show correlated LLM trading behaviour in a single-asset simulation, listing multi-asset settings, endogenous liquidity and decorrelation mechanisms as open problems; and \citet{nagaraj2026} argue, in a purely mathematical treatment, that risk from shared foundation models cannot be diversified away and must be prevented ex ante. \citet{danielsson2025} show in a game-theoretic model that AI makes financial crises faster and more severe. Supervisors have flagged herding and concentration as channels through which AI could threaten stability \citep{fsb2024,imf2024,boe2025,iosco2025,gensler2020,danielsson2024}. What is missing is a model that links model concentration to synchronized liquidation through the actual mechanics of a wealth platform, a fragility measure calibrated to losses that supervisors could compute, and evidence on which interventions work and at what cost to clients.

\paragraph{Contributions.} This paper provides those elements.
\begin{enumerate}[leftmargin=1.5em,itemsep=2pt]
\item \textbf{A model of the agentic wealth layer.} We build a multi-asset agent-based model in which each client's agent runs the full pipeline---perception, tactical allocation, automated risk check and execution---and aggregate order flow moves prices through a latent-liquidity version of the square-root impact law \citep{toth2011,bouchaud2018} with endogenous volatility. Fundamentals are block-bootstrapped from \RealDays{} days of real S\&P~500 constituent returns (1990--2022). Three correlation channels are explicit: shared data and risk models (C1), model monoculture (C2) and synchronized automation (C3).
\item \textbf{Theory.} We show that the effective number of independent decision-makers in the layer is bounded no matter how many clients it serves---\NeffLimitConsumer{} at the measured consumer concentration (Proposition~\ref{prop:floor})---derive the synchronized-liquidation multiplier and a critical agentic share under linear impact (Proposition~\ref{prop:mult}), prove that correlation can be measured from secure-aggregated sums with a privacy cost that vanishes as the client population grows (Proposition~\ref{prop:private}), and characterise what dispersing execution can do against volatility- and loss-triggered de-risking when agents react to their own impact (Proposition~\ref{prop:dither}). Each result is verified by Monte Carlo, and the full model is used to test where the stylised predictions hold and where they fail.
\item \textbf{A calibrated fragility index.} We construct an index from quantities a supervisor can observe---aggregate agentic flow, obtainable under secure aggregation, and public returns---and calibrate it to tail liquidation in basis points out of sample across \EtwoN{} simulated economies.
\item \textbf{A mitigation frontier.} We evaluate six interventions and two combinations in exactly paired simulations, scoring each by tail reduction against client welfare cost.
\item \textbf{Privacy-preserving measurement and control.} We show how the index and a participation circuit breaker can run on differentially private aggregates, and quantify when privacy is effectively free.
\end{enumerate}

\FindingsParagraph

The paper is organised as follows. \Cref{sec:lit} reviews related work. \Cref{sec:model} presents the model, \cref{sec:theory} the theory and \cref{sec:calib} the calibration. \Cref{sec:results} reports how fragility depends on adoption, concentration and architecture, including misinformation shocks and historical counterfactuals. \Cref{sec:index} builds the fragility index, \cref{sec:mitig} evaluates mitigations, and \cref{sec:privacy} treats privacy-preserving measurement. \Cref{sec:policy} discusses policy, \cref{sec:limits} limitations, and \cref{sec:concl} concludes.

\section{Related literature}\label{sec:lit}

\paragraph{Endogenous risk, fire sales and liquidity spirals.} A long literature shows that common risk-management rules turn exogenous shocks into endogenous ones. \citet{danielsson2004} show that value-at-risk constraints applied uniformly make prices more volatile; \citet{brunnermeier2009} formalise the interaction between market and funding liquidity; \citet{contwagalath2013} show how distressed selling creates endogenous correlation; \citet{greenwood2015} and \citet{caccioli2014} model contagion through overlapping portfolios; and \citet{pedersen2009} documents crowded exits. Historical episodes---portfolio insurance in 1987 \citep{brady1988,gennotte1990,jacklin1992}, the 2007 quant unwind \citep{khandani2011} and the 2010 Flash Crash \citep{kirilenko2017}---ground these mechanisms. Volatility-managed portfolios \citep{moreira2017,harvey2018} and their behaviour in sell-offs \citep{bis2018,ecb2020} motivate the overlay channel. We contribute a setting in which the source of common behaviour is not overlapping holdings or a regulatory rule but a shared foundation model, and in which the agents are retail wealth clients rather than leveraged institutions.

\paragraph{Algorithmic monoculture and AI in finance.} \citet{kleinberg2021} show that a common algorithm can lower welfare even when it is individually more accurate, and \citet{bommasani2021,bommasani2022} document outcome homogenisation from shared foundation models. In finance, \citet{gensler2020} warn that uniform deep-learning models may increase herding; \citet{danielsson2024} discuss AI in financial regulation; \citet{danielsson2025} show that common data, speed and strategic complementarities make AI-driven crises faster and more severe; and supervisory reports identify concentration and correlation as stability channels \citep{fsb2024,imf2024,boe2025,iosco2025}. The closest papers are \citet{frimpong2026}, who proposes a qualitative monoculture index; \citet{sahoo2026}, who calibrate a normalised error-correlation index for credit; \citet{ng2026}, who classifies AI-driven flash-crash mechanisms; \citet{ross2026}, who show correlated LLM trading behaviour and a non-diversifiable risk floor in a single-asset market; and \citet{nagaraj2026}, who argue mathematically that systematic risk from shared foundation models is non-diversifiable and call for structural prevention. Our results qualify the last conclusion: diversifying providers does not reduce market-wide liquidation in our model, but decorrelating automation does, so structural prevention should target synchronized rules rather than model choice alone. Relative to these papers we model the wealth-management layer explicitly, add multi-asset trading, latent liquidity, automated overlays and attention-driven execution, calibrate an index to losses, test mitigations, and provide a privacy-preserving measurement protocol.

\paragraph{LLM agents in markets.} \citet{lopezlira2025} studies LLM agents trading in simulated markets; \citet{chen2024} document extrapolation and miscalibration in LLM return forecasts; \citet{xia2026} survey agentic trading; \citet{na2026} show role-specific poisoning attacks on multi-agent trading systems; \citet{yan2025} show that LLM trading agents' stated reasoning is not reliably faithful; and \citet{wang2026} audit robustness and security failures of academic LLM trading schemes. \citet{kutasov2025} evaluate control protocols for untrusted agents. These papers concern individual agents or trading firms; we study the aggregate consequences of millions of client-facing agents built on the same few models.

\paragraph{Robo-advice, microstructure and privacy.} Robo-advice improves diversification for many investors \citep{dacunto2019}; retail turnover and attention shape trading volume \citep{barber2000,barber2008}. Our market block builds on the square-root impact law and latent-liquidity models \citep{kyle1985,toth2011,bouchaud2018}, optimal execution \citep{almgren2001}, liquidity provision in stress \citep{nagel2012} and circuit breakers \citep{greenwald1991,subrahmanyam1994}. The measurement protocol uses secure aggregation \citep{bonawitz2017} and differential privacy with R\'enyi accounting \citep{dwork2014,mironov2017,mcmahan2018,canonne2020}; the effective-number measure follows the design-effect and diversification literatures \citep{kish1965,meucci2009}. Methodologically we follow the agent-based tradition in financial-stability analysis \citep{farmer2009,bookstaber2018}.

\section{A model of the agentic wealth layer}\label{sec:model}

\begin{figure}[t]
\centering
\includegraphics[width=\textwidth]{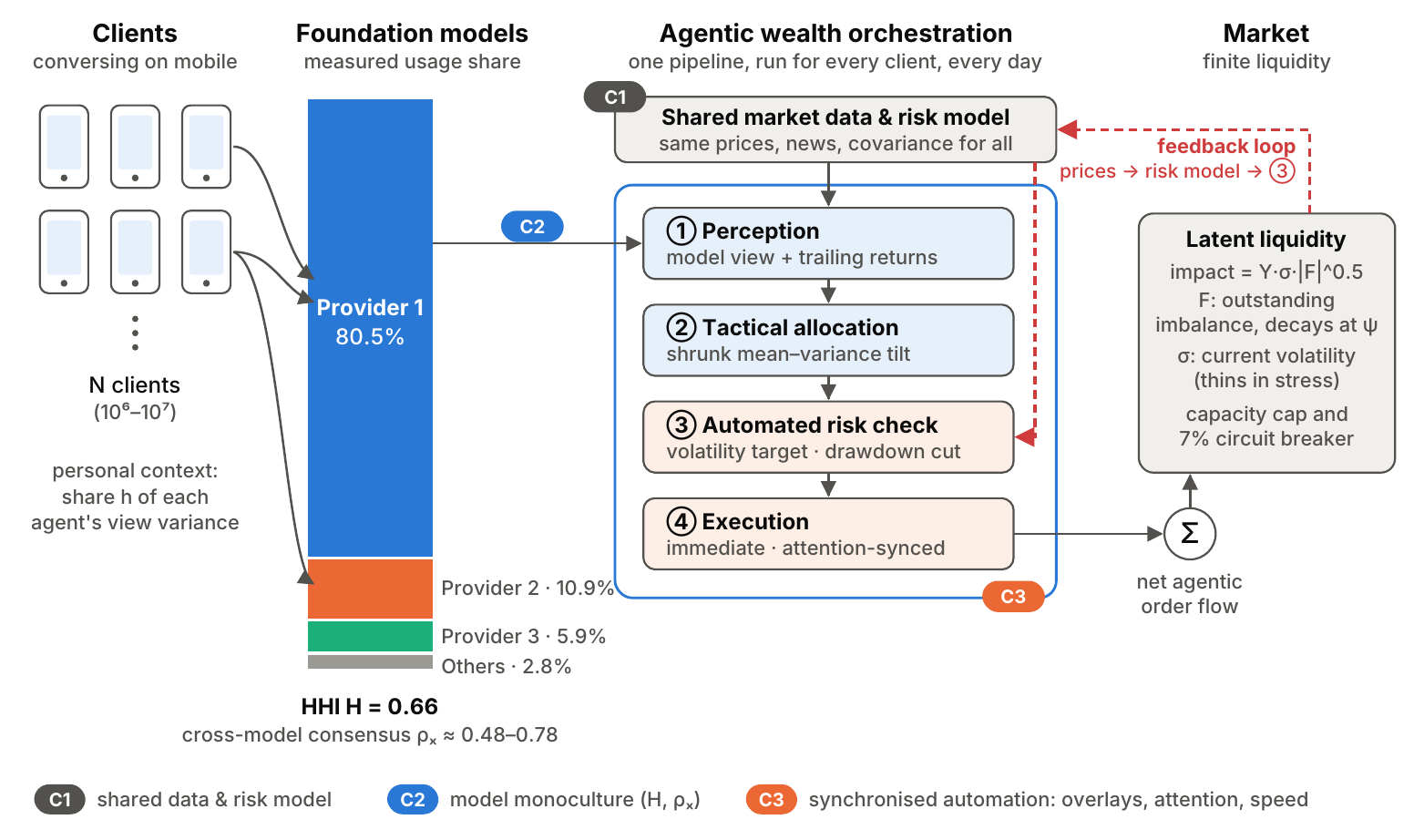}
\caption{The agentic wealth-management layer. Clients converse with agents built on a few foundation models (shares measured by \citealp{statcounter2026}). Every agent runs the same four-step pipeline on shared market data and risk models; aggregate net order flow moves prices through finite latent liquidity, and realised prices feed back into the risk check. Badges mark the three correlation channels.}
\label{fig:arch}
\end{figure}

\Cref{fig:arch} summarises the model. Time is discrete (trading days). There are $K$ risky assets and cash, $N$ client agents, and $M$ foundation-model providers.

\subsection{Fundamentals and market}
Fundamental log returns $r^f_t\in\mathbb{R}^K$ are drawn by a stationary block bootstrap \citep{politis1994} (mean block length 20 days) from the daily returns of the $K=20$ S\&P~500 constituents bundled with the open-source \texttt{skfolio} library \citep{skfolio}, 1990--2022. The bootstrap preserves fat tails, volatility clustering and the rise of cross-asset correlation in stress. In historical counterfactuals (\cref{sec:hist}) we instead replay actual paths. Assets have equal market capitalisation; daily dollar volume in each asset is a fraction $\tau$ of its capitalisation.

Agentic order flow in asset $k$, $q_{k,t}$, is measured in units of that asset's daily volume. Following latent-liquidity models of the square-root law \citep{toth2011,bouchaud2018}, impact is a concave function of the outstanding, not-yet-absorbed imbalance $F_{k,t}$, which decays as liquidity refills:
\begin{equation}\label{eq:impact}
F_{k,t} = (1-\psi)F_{k,t-1} + q_{k,t},\qquad
d_{k,t} = Y\,\hat\sigma_{k,t}\,\operatorname{sign}(F_{k,t})\,|F_{k,t}|^{\delta},\qquad
r_{k,t} = r^f_{k,t} + d_{k,t}-d_{k,t-1},
\end{equation}
where $d_{k,t}$ is the agent-induced log price deviation, $\delta=1/2$, $Y$ is the impact prefactor and $\hat\sigma_{k,t}$ is the current EWMA volatility. Because $\hat\sigma$ is estimated from realised returns that include agent impact, liquidity thins endogenously in stress \citep{nagel2012}. For a fast metaorder \eqref{eq:impact} reproduces the square-root law; unlike applying the law to each day's flow separately, it does not penalise splitting an order over several days, and slower execution lowers peak impact because $F$ decays. Agentic flow is capped at one day's volume per asset, and a 7\% market-wide circuit breaker \citep{nasdaqmwcb} halts further agent-driven decline on a day. The case $\delta=1$ (linear impact) is used as a theoretical benchmark.

\subsection{Agents and perception}
Agent $i$ is assigned to provider $m(i)$ with probabilities $s_m$; it has a wealth weight, a suitability bucket that fixes its strategic equity share $E_i\in\{0.30,0.45,0.60,0.75,0.90\}$ and risk aversion $\gamma_i$. Each day the agent forms expected excess returns
\begin{equation}\label{eq:view}
\mu_{i,t} = \underbrace{s_v\Big[\sqrt{1-h}\,\big(\sqrt{\rho_x}\,\beta_t + \sqrt{1-\rho_x}\,\eta_{m(i),t}\big) + \sqrt{h}\,u_{i,t}\Big]}_{\text{non-corrective view}} \;+\; g\,\bar r_t ,
\end{equation}
where $\beta_t$ is a cross-model consensus shock common to all providers, $\eta_{m,t}$ a provider-specific interpretation shock, $u_{i,t}$ a client-specific component reflecting personal context, and $\bar r_t$ an exponentially weighted trailing return from the shared feed. All three shocks are independent persistent AR(1) processes with unit variance. The parameter $h$ is the personalisation share of view variance; $\rho_x$ is the share of model-level variance that is common across providers; and $g$ captures extrapolation \citep{chen2024}.

\subsection{Allocation, risk check and execution}
Agents tilt around a strategic allocation $E_i w^{\mathrm{SAA}}$ with equal weights:
$w^\star_{i,t} = \Pi\big(E_i w^{\mathrm{SAA}} + \theta\,\tilde\Sigma_t^{-1}\mu_{i,t}\big)$,
where $\tilde\Sigma_t$ is the shared EWMA covariance ($\lambda=0.94$; \citealp{riskmetrics1996}) shrunk halfway to its diagonal \citep{ledoit2004}, $\theta$ is set so that the typical tactical tracking error is \num{0.8}\%, and $\Pi$ imposes long-only positions, a 25\% position cap and a budget constraint. The tactical target is refreshed only when the client converses with the agent; the probability of a conversation is $p_t = \min\{1, p_0 + p_1 |r^{\mathrm{idx}}_{t-1}|/\sigma^{\mathrm{idx}}\}$, so activity rises after large market moves \citep{barber2008}.

The automated risk check runs every day for every agent. A fraction $f_{vt}$ of agents applies a volatility target, scaling risky exposure by $\ell_{i,t}=\min\{1,\sigma^\star_i/\hat\sigma^p_{i,t}\}$ with $\sigma^\star_i=0.2E_i$ and a 5-point rebalancing threshold; a fraction $f_{dd}$ applies a drawdown control that halves exposure when the client's drawdown exceeds 15\% and restores it when the drawdown recovers to half that level. Trades close the gap between holdings and $\ell_{i,t}w^\star_{i,t}$ when it exceeds a 0.5-point band per asset, spread over an execution horizon of $H_{\mathrm{exec}}$ days ($H_{\mathrm{exec}}=1$ in the agentic baseline). Agents pay half the day's impact on traded notional.

\paragraph{Traditional benchmark.} To isolate architecture, the \emph{traditional} robo-adviser uses identical views but reviews each client once every 21 trading days on a staggered schedule, applies the risk check only at reviews, rebalances only when portfolio drift exceeds 3\%, and ignores attention.

\subsection{Outcome measures}
The agent-induced index deviation is $d^{\mathrm{idx}}_t = K^{-1}\sum_k d_{k,t}$. Our headline tail measure is the \emph{synchronized liquidation at risk},
\begin{equation}\label{eq:slar}
\SLaR_{99} = -Q_{0.01}\big(d^{\mathrm{idx}}_{t+5}-d^{\mathrm{idx}}_{t}\big)\times 10^4 ,
\end{equation}
the 1\% quantile of five-day agent-induced index moves, in basis points. Because $d$ isolates the agentic component, $\SLaR$ is zero without agents irrespective of fundamentals. Client welfare is the wealth-weighted annualised certainty equivalent $\mathrm{CE}=\sum_i \omega_i(\bar r_i - \tfrac{\gamma_i}{2}\hat\sigma_i^2)$ of realised portfolio returns net of impact costs. We also report maximum agent-induced deviation, excess volatility relative to fundamentals, turnover and costs. All randomness is split into independent streams (fundamentals, agent attributes, views, conversations, mitigation noise), so runs that differ only in a mitigation share every fundamental path, client and view exactly.

\section{Theory}\label{sec:theory}

We state four results in a linearised version of the model; proofs are in \cref{app:proofs} and each result is checked by Monte Carlo (\cref{tab:theory}).

\subsection{The monoculture floor}
Let $n_m$ agents use provider $m$, $N=\sum_m n_m$, $H_N=\sum_m (n_m/N)^2$ and $\tilde H=\sum_m n_m(n_m-1)/[N(N-1)]$. Consider the non-corrective views in \eqref{eq:view} and linear demands $x_i = a\, v_i$.

\begin{proposition}[Monoculture floor]\label{prop:floor}
(i) The mean pairwise correlation of agent views is $\bar\rho = (1-h)\big[\rho_x + (1-\rho_x)\tilde H\big]$.
(ii) The variance of per-capita demand is $\Var(\bar x)=a^2 s_v^2\big[(1-h)C_N + h/N\big]$ with $C_N=\rho_x+(1-\rho_x)H_N$.
(iii) With shares fixed and $N\to\infty$, the effective number of independent decision-makers $\Neff = N/[1+(N-1)\bar\rho]$ converges to $1/[(1-h)C]$, where $C=\rho_x+(1-\rho_x)H$ is the \emph{effective concentration}.
(iv) Provider diversification ($H\to 0$) cannot reduce aggregate demand variance below $a^2s_v^2(1-h)\rho_x$.
\end{proposition}

At the measured consumer concentration ($H=\IntroHHIConsumer$), $\rho_x=0.48$ and $h=0.4$, the layer behaves like $\NeffLimitConsumer$ independent investors however many clients it serves; at enterprise concentration the figure is $\NeffLimitEnterprise$; and even perfect provider diversity raises it only to $\NeffLimitDiverse$. Part (iv) generalises the risk floor of \citet{ross2026} to a market with many providers and unequal shares.

\subsection{The synchronized-liquidation multiplier}
\begin{proposition}[Multiplier and critical share]\label{prop:mult}
Let impact be linear, $\Delta p=\lambda Q/D$, and let a fraction $\pi$ of agentic wealth (a share $\alpha$ of the market) run an overlay whose net trade per unit price move is $\varepsilon$; for a loss-linear rule with multiplier $m$ and risky exposure $e$, $\varepsilon=(m-1)e$. An exogenous move $\Delta p_0$ cumulates to $\Delta p_0/(1-\kappa)$ with loop gain $\kappa=\pi\varepsilon\lambda\alpha/D$. The cascade is finite iff $\kappa<1$, which defines a critical agentic share $\alpha^\star=D/(\pi\varepsilon\lambda)$.
\end{proposition}

The term $-1$ in $\varepsilon$ reflects that holdings drift passively with prices; omitting it overstates the gain. Under the concave impact of \eqref{eq:impact} the local gain $\kappa(F)\propto|F|^{\delta-1}$ falls as the imbalance grows, so cascades are self-limiting rather than explosive, while the gain rises with $\hat\sigma$; fragility is therefore state-dependent. Combining both propositions motivates an \emph{amplification law} for tail liquidation,
\begin{equation}\label{eq:law}
\SLaR \;\propto\; \frac{Y\,\sigma\,\big[\mathrm{sd}(q)\big]^{\delta}}{1-\kappa},\qquad \mathrm{sd}(q)\;\approx\;\frac{\alpha}{\tau}\,\phi\,\sqrt{\bar\rho_f},
\end{equation}
where $\phi$ is the gross turnover rate of agentic portfolios and $\bar\rho_f$ the mean pairwise correlation of agent flows. Monoculture enters through $\bar\rho_f$ (shock size); automation enters through $\kappa$ (propagation). \Cref{sec:index} turns \eqref{eq:law} into a measurable index.

\subsection{Measuring correlation from sums}
\begin{proposition}[Private measurement]\label{prop:private}
Let each agent standardise its own flow, $z_{i,t}=f_{i,t}/s_i$, with $\E z=0$, $\Var z=1$, and let $S_t=\sum_i z_{i,t}$. Then $\E[S_t^2]=N+N(N-1)\bar\rho$, so $\Neff=N^2/\E[S_t^2]$. If only $\tilde S_t = S_t+\nu_t$ is released, with $\nu_t\sim\mathcal N(0,\sigma_\nu^2)$, then $\hat N_{\mathrm{eff}} = N^2/(T^{-1}\sum_t\tilde S_t^2-\sigma_\nu^2)$ is consistent, and for Gaussian, serially independent $S_t$ its relative standard error is approximately $\sqrt{2/T}\,(1+\sigma_\nu^2\Neff/N^2)$. With $|z|\le c$ and $\sigma_\nu=zc$, the $T$ releases satisfy client-level $(\varepsilon,\delta)$-differential privacy with $\varepsilon=\min_{a>1}\{Ta/(2z^2)+\log\frac{a-1}{a}-\frac{\log\delta+\log a}{a-1}\}$.
\end{proposition}

Because $S_t$ is a sum, it can be computed by secure aggregation across platforms without revealing any platform's or client's flows \citep{bonawitz2017}. The privacy penalty $\sigma_\nu^2\Neff/N^2$ vanishes as the client population grows, because the correlated signal scales with $N^2$ while client-level noise does not.

\subsection{What dispersing execution can and cannot do}
\begin{proposition}[Execution dispersion]\label{prop:dither}
Let a liquidation of size $X$ be executed evenly over $H$ periods with linear impact $\lambda$ and an EWMA variance update with weight $\vartheta$. (i) The variance estimate rises by $\vartheta\lambda^2X^2/H$, so to first order volatility-triggered de-risking falls by the factor $1/H$. (ii) The cumulative loss $\lambda X$, which drives loss-triggered de-risking, does not depend on $H$. (iii) With impact decay $\psi$, peak deviation falls by $g(H)=[1-(1-\psi)^H]/(\psi H)$.
\end{proposition}

Proposition~\ref{prop:dither} isolates the agent's reaction to its \emph{own} impact under linear, permanent impact. Read naively, it says that slower execution damps volatility-target spirals but leaves drawdown spirals intact. Two of its assumptions need not hold in practice: volatility targets may be triggered mainly by common \emph{fundamental} volatility seen through a shared risk model, which execution speed cannot change; and with impact decay and volatility-dependent liquidity, spreading a large exposure cut does reduce the cumulative loss it causes. \Cref{sec:prop4test} tests the prediction in the full model and finds the assignment of levers reversed.

\begin{table}[t]\centering\small
\caption{Monte Carlo verification of the propositions. Proposition~\ref{prop:floor}: $N=1{,}500$ agents, $T=4{,}000$ periods, nine combinations of concentration ($H$), cross-model consensus ($\rho_x$) and personalisation ($h$). Proposition~\ref{prop:mult}: iterated rounds of a loss-linear overlay with linear impact. Proposition~\ref{prop:private}: $N=2{,}000$, $T=750$, 300 replications. Proposition~\ref{prop:dither}: exact algebra.}
\label{tab:theory}
\begin{tabular}{llrr}\toprule
Result & Quantity & Theory & Monte Carlo \\ \midrule
Prop.~\ref{prop:floor} & largest rel.\ error, $\Var(\bar x)$ (9 cases) & -- & 3.5\% \\
 & largest $|\bar\rho_{\mathrm{MC}}-\bar\rho|$ (9 cases) & -- & 0.019 \\
 & $\Neff$, $H=0.66$, $\rho_x=0.48$, $h=0.4$ & 2.01 & 1.98 \\
 & $\Neff$, $H=0.28$, $\rho_x=0.48$, $h=0.4$ & 2.65 & 2.68 \\
 & $\Neff$, $H=0.20$, $\rho_x=0.48$, $h=0.4$ & 2.85 & 2.94 \\
Prop.~\ref{prop:mult} & cumulative move$/\Delta p_0$, $\kappa=0.20$ & 1.25 & 1.25 \\
Prop.~\ref{prop:mult} & cumulative move$/\Delta p_0$, $\kappa=0.50$ & 2.00 & 2.00 \\
Prop.~\ref{prop:mult} & cumulative move$/\Delta p_0$, $\kappa=0.80$ & 5.00 & 5.00 \\
Prop.~\ref{prop:mult} & cumulative move$/\Delta p_0$, $\kappa=0.95$ & 20.00 & 19.89 \\
Prop.~\ref{prop:private} & rel.\ s.e.\ of $\hat N_{\mathrm{eff}}$, $\bar\rho=0.3$, $\sigma_\nu=0$ & 0.052 & 0.050 \\
Prop.~\ref{prop:private} & rel.\ s.e.\ of $\hat N_{\mathrm{eff}}$, $\bar\rho=0.3$, $\sigma_\nu=50$ & 0.052 & 0.051 \\
Prop.~\ref{prop:private} & rel.\ s.e.\ of $\hat N_{\mathrm{eff}}$, $\bar\rho=0.05$, $\sigma_\nu=50$ & 0.052 & 0.054 \\
Prop.~\ref{prop:private} & rel.\ s.e.\ of $\hat N_{\mathrm{eff}}$, $\bar\rho=0.3$, $\sigma_\nu=500$ & 0.062 & 0.064 \\
Prop.~\ref{prop:dither} & vol.-trigger input ratio, $H=1$ & 1.00 & 1.00 \\
Prop.~\ref{prop:dither} & vol.-trigger input ratio, $H=2$ & 0.50 & 0.50 \\
Prop.~\ref{prop:dither} & vol.-trigger input ratio, $H=5$ & 0.20 & 0.20 \\
Prop.~\ref{prop:dither} & vol.-trigger input ratio, $H=10$ & 0.10 & 0.10 \\
\bottomrule\end{tabular}\end{table}

\section{Calibration}\label{sec:calib}
\begin{table}[t]
\centering\small
\caption{Calibration. Baseline values; ranges are explored in robustness checks and in the ensemble of \cref{sec:index}.}
\label{tab:calib}
\begin{tabularx}{\textwidth}{>{\raggedright\arraybackslash}p{4.3cm}>{\raggedright\arraybackslash}p{3.2cm}X}
\toprule
Parameter & Baseline & Source / target \\
\midrule
\multicolumn{3}{l}{\emph{Market}}\\
Fundamental returns & bootstrap, block 20 d & 20 S\&P~500 constituents, 1990--2022, \RealDays{} days \citep{skfolio,politis1994} \\
Market turnover $\tau$ & 100\%/yr & U.S. value traded/capitalisation, 68--165\% in 2015--2019 \citep{worldbank2019} \\
Impact law & $Y=1$, $\delta=0.5$ & square-root law \citep{toth2011,bouchaud2018} \\
Impact decay $\psi$ & 0.05/day & latent-liquidity refill; 0.02--0.15 in robustness \\
Circuit breaker & 7\% index decline & market-wide circuit breaker \citep{nasdaqmwcb} \\
\multicolumn{3}{l}{\emph{Model concentration and correlation}}\\
Provider shares (consumer) & 80.5/10.9/5.9/1.8/1.0\% & AI-chatbot usage, Sep.~2026; $H=\IntroHHIConsumer$ \citep{statcounter2026} \\
Provider shares (enterprise) & 40/27/21/12\% & enterprise API spend, Dec.~2025; $H\approx\IntroHHIEnterprise$ \citep{menlo2025} \\
Cross-model consensus $\rho_x$ & 0.48 (stress 0.78) & risk floor $\approx$ half single-model variance \citep{ross2026}; inter-model sentiment correlation \citep{chen2024} \\
Personalisation share $h$ & 0.4 & 0.2--0.9 in the ensemble \\
Extrapolation $g$ & trailing-return term = half the view s.d. & LLM forecasts extrapolate $\approx3\times$ more than humans \citep{chen2024}; 0--3$\times$ tested \\
\multicolumn{3}{l}{\emph{Agents}}\\
Agentic share $\alpha$ & 5\% (1--15\%) & adoption survey \citep{investingcom2026}; largest digital-advice platforms \citep{investingintheweb2026}; volatility strategies up to \$2tn \citep{ecb2020} \\
Tactical tracking error & 0.8\% (3\% high conviction) & annual turnover \TurnoverAgentic{}, cf.\ 75\% for households \citep{barber2000} \\
Risk model & EWMA $\lambda=0.94$, 50\% shrinkage & \citet{riskmetrics1996}; \citet{ledoit2004} \\
Volatility-target overlay & $f_{vt}=0.5$; $\sigma^\star=0.2E_i$ & volatility-managed portfolios \citep{moreira2017,harvey2018} \\
Drawdown overlay & $f_{dd}=0.3$; 15\% trigger, 50\% cut & loss-linear rules \citep{gennotte1990,brady1988} \\
Conversations & $p_0=0.1$, $p_1=0.15$ & attention-driven activity \citep{barber2008} \\
Execution & immediate; 0.5-pt band & agentic baseline; traditional robo: 21-day staggered reviews, 3\% drift band \\
\bottomrule
\end{tabularx}
\end{table}

\Cref{tab:calib} lists parameters and sources. Three choices deserve comment. First, model concentration uses measured usage shares in two channels: consumer chatbots, through which retail clients increasingly obtain advice \citep{statcounter2026}, and enterprise APIs, on which wealth platforms build \citep{menlo2025}. Underlying-model concentration is likely higher than the consumer figure because several consumer products run on the same underlying models; our measured $H$ is therefore conservative. Second, cross-model consensus has two independent anchors: \citet{ross2026} estimate a correlation floor of roughly half the single-model variance, $\rho_x\approx0.48$, and \citet{chen2024} report a correlation of $0.78$ between two leading models' sentiment measures; we use the former as baseline and the latter as a stress case. Third, we calibrate tactical activity so that agentic portfolios turn over \TurnoverAgentic{} times a year, close to the 75\% household turnover documented by \citet{barber2000}, rather than letting view noise generate unrealistic churn. The agentic share $\alpha$ is a scenario variable between 1\% and 15\% of market capitalisation, anchored by the adoption survey above, the size of the largest digital advice platforms (one reports \$344.6~billion under advice; \citealp{investingintheweb2026}) and the up to \$2~trillion in volatility strategies \citep{ecb2020}.

\Cref{tab:facts} compares the model with the data. Without agents the bootstrap reproduces the volatility, fat tails, volatility clustering and the rise of cross-asset correlation in stress of the real constituents (index volatility \FactsVolNone\% against \RealIdxVol\%; correlation \RealCorrCalm{} in calm and \RealCorrStress{} in stressed markets in the data). Adding the agentic layer leaves these unconditional statistics nearly unchanged, even at $\alpha=15\%$: the layer's effect is concentrated in the tail of short-horizon moves, which is why we measure it with $\SLaR$ rather than with volatility.\begin{table}[t]\centering\small
\caption{Stylised facts: real data and the model. Equal-weighted index of the 20 constituents. Calm and stress days are the bottom half and top decile of 21-day rolling index volatility. Model columns use one ten-year bootstrap economy without agents and with the agentic baseline at $\alpha=5\%$ and $\alpha=15\%$.}
\label{tab:facts}
\begin{tabular}{lrrrr}\toprule
Statistic & Real 1990--2022 & No agents & $\alpha=5\%$ & $\alpha=15\%$ \\ \midrule
Annualised index volatility (\%) & 18.9 & 19.5 & 19.5 & 19.7 \\
Excess kurtosis of index returns & 9.4 & 8.5 & 7.7 & 7.6 \\
Mean pairwise correlation, calm & 0.18 & 0.16 & 0.16 & 0.15 \\
Mean pairwise correlation, stress & 0.48 & 0.52 & 0.50 & 0.50 \\
Lag-1 autocorrelation of $|r|$ & 0.28 & 0.26 & 0.28 & 0.30 \\
Lag-1 autocorrelation of $r$ & -0.071 & -0.059 & -0.080 & -0.089 \\
\bottomrule\end{tabular}\end{table}

\section{How fragility depends on adoption, architecture and concentration}\label{sec:results}

\subsection{Adoption and architecture}
\begin{figure}[t]
\centering
\includegraphics[width=\textwidth]{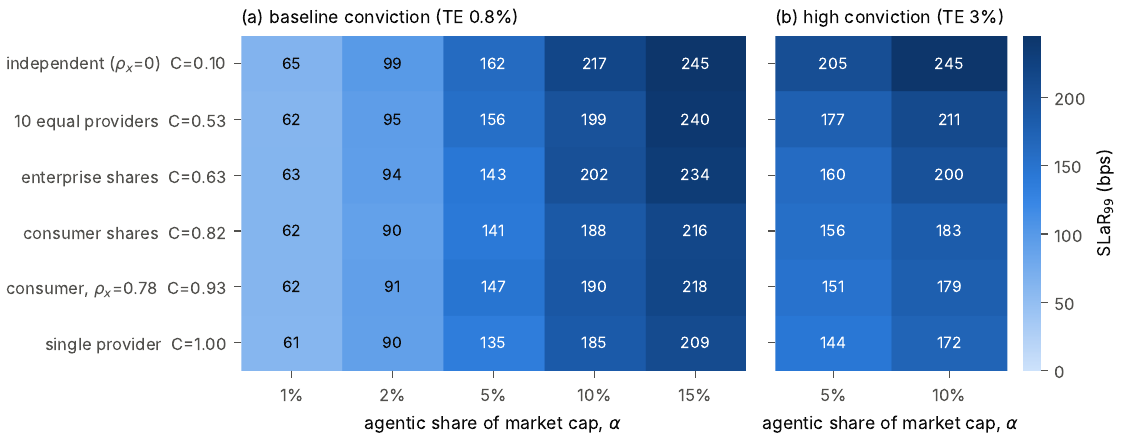}
\caption{Tail liquidation ($\SLaR_{99}$, bps) by agentic share and model-concentration scenario, mean of three seeds. Scenarios are ordered by effective concentration $C=\rho_x+(1-\rho_x)H$. (a) Baseline tactical conviction (0.8\% tracking error). (b) High conviction (3\%).}
\label{fig:phase}
\end{figure}

\Cref{fig:phase}a reports $\SLaR_{99}$ across agentic shares and model-concentration scenarios. Tail liquidation rises steadily with adoption: at consumer concentration it grows from \EoneSLaRConsOne{}~bps at $\alpha=1\%$ to \EoneSLaRConsFive{}~bps at 5\% and \EoneSLaRConsFifteen{}~bps at 15\%. A regression of $\log\SLaR_{99}$ on $\log\alpha$ and effective concentration across all baseline-conviction economies gives an elasticity with respect to $\alpha$ of \EoneElasAlpha{} ($\pm$\EoneElasAlphaSE{} s.e.), close to the one-half implied by the square-root impact law and by \eqref{eq:law}. Within the calibrated range there is no critical share at which the layer becomes explosive: the concavity of impact makes cascades self-limiting (Proposition~\ref{prop:mult}). Under linear impact and universal overlays, by contrast, $\SLaR_{99}$ rises from \EeightLinLo{} to \EeightLinHi{}~bps as $\alpha$ goes from 5\% to 20\%, the estimated loop gain reaches \EeightLinKappaHi{}, and the market-wide circuit breaker begins to bind (on \EeightLinMwcbHi\% of days); the critical-share mechanism is present but held in check by concave impact.

\begin{figure}[t]
\centering
\includegraphics[width=0.6\textwidth]{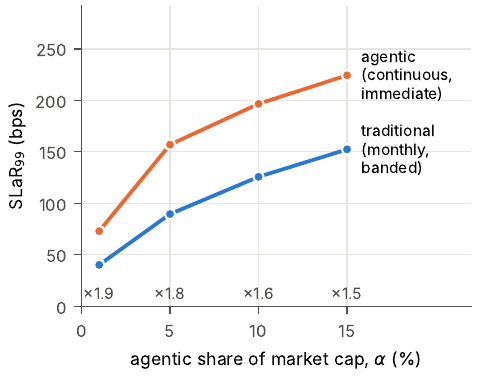}
\caption{Agentic versus traditional architecture with identical views. Mean $\SLaR_{99}$ over three seeds; numbers along the bottom give the paired ratio of agentic to traditional at each adoption level.}
\label{fig:arch2}
\end{figure}

Architecture matters as much as adoption. Holding views, overlays and turnover-generating preferences fixed, the agentic architecture---continuous risk checks, immediate execution and attention-driven conversations---raises $\SLaR_{99}$ relative to a traditional robo-adviser that reviews clients monthly on a staggered schedule (\cref{fig:arch2}). The paired ratio is \EarchRatioFive{} ($\pm$\EarchRatioFiveCI) at $\alpha=5\%$ and lies between \EarchRatioLo{} and \EarchRatioHi{} across adoption levels: the same advice, delivered at the speed of a chat, produces roughly one and a half to two times the tail liquidation.

The agentic layer is nearly invisible in unconditional statistics. In a ten-year economy, index volatility is \FactsVolNone\% without agents and \FactsVolFifteen\% with an agentic share of 15\%, and kurtosis does not rise (\cref{tab:facts}). The risk sits in the tail and in its timing: a supervisor monitoring volatility would not see it.

\subsection{Model concentration: a driver of crowding, not of index liquidation}\label{sec:concentration}
Proposition~\ref{prop:floor} implies that, at measured concentration, the agentic layer acts like about \NeffLimitConsumer{} independent investors in its \emph{views}. One might therefore expect index-level tail risk to rise with concentration. It does not (\cref{fig:phase}a). Across scenarios from fully independent views ($C=0.10$) to a single provider ($C=1$), $\SLaR_{99}$ changes little and, if anything, falls: the regression coefficient on $C$ is \EoneCoefC{} ($\pm$\EoneCoefCSE{}) at baseline conviction and \EoneHCCoefC{} ($\pm$\EoneHCCoefCSE{}) when agents hold high-conviction views (\cref{fig:phase}b), where independent views produce \EoneHCIndepFive{}~bps at $\alpha=5\%$ against \EoneHCSingleFive{}~bps for a single provider.

The mechanism is the one identified in \cref{sec:twokinds}. Views mostly rotate money between assets, so concentrated views create large, offsetting asset-level imbalances. Under concave impact these imbalances absorb part of the common de-risking flow generated by overlays and drawdown controls; when views are independent, rotations net out and common selling meets fresh liquidity. \ELinText{} Concentration therefore matters---it drives crowding in individual securities (\cref{sec:twokinds}) and it determines how far a shared error propagates (\cref{sec:misinfo})---but market-wide liquidation is governed by synchronized automation.

\subsection{Shared misinformation: where concentration matters directly}\label{sec:misinfo}
\begin{figure}[t]
\centering
\includegraphics[width=\textwidth]{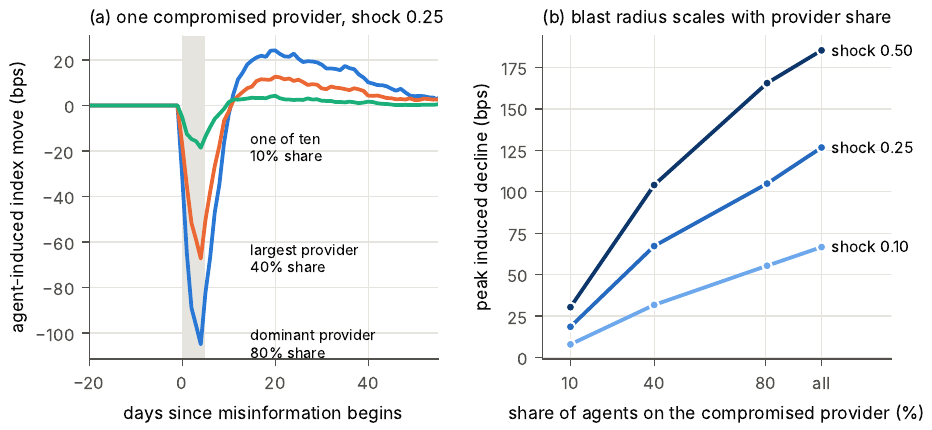}
\caption{Shared misinformation. (a) Average agent-induced index move after a five-day shock (shaded) to the views of one provider, relative to the identical economy without the shock. (b) Peak induced decline against the share of agents on the compromised provider; ``all'' is a shock to every provider. Four seeds per point.}
\label{fig:misinfo}
\end{figure}

Concentration becomes decisive when a single error reaches many agents at once. We inject a five-day misinformation shock into agents' expected returns---a pessimistic reading of the news, a poisoned retrieval source or a compromised data feed, of the kind documented for multi-agent trading systems by \citet{na2026}---either into the consensus component, so that every provider receives it, or into the views of one provider only (\cref{fig:misinfo}). Each shocked economy is compared with the identical economy without the shock.

When every provider is affected, the induced decline is the same whatever the concentration---about \EsixConsCons{}~bps for a shock of size 0.25 under consumer shares and \EsixDivCons{}~bps with ten equal providers---because every agent receives the same signal. When one provider is compromised, the damage scales with its share. The same shock to the dominant provider under consumer shares (80\% of agents) produces a peak decline of \EsixConsMod{}~bps; to the largest enterprise provider (40\%), \EsixEntMod{}~bps; to one of ten equal providers, \EsixDivMod{}~bps. Concentration multiplies the blast radius of a single-provider failure by about \EsixModRatio{} relative to a diversified market. Automation-level mitigation helps here too: with M7 in place, the induced decline from a shock to all providers falls from \EsixConsCons{} to \EsixConsMSeven{}~bps. Model diversity is thus a defence against \emph{idiosyncratic} model failure, while automation-level design is a defence against \emph{common} shocks and against the liquidation dynamics that follow either kind.

\subsection{Historical counterfactuals}\label{sec:hist}
\begin{figure}[t]
\centering
\includegraphics[width=\textwidth]{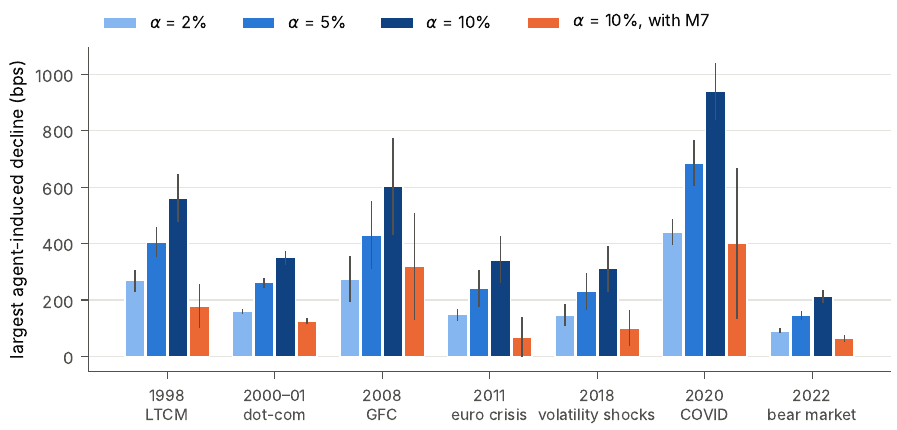}
\caption{Historical counterfactuals: largest agent-induced decline during each episode when real returns are replayed with the agentic layer added (three seeds, 95\% intervals).}
\label{fig:hist}
\end{figure}

To connect the model with episodes that supervisors know, we replay the actual daily constituent returns of seven stress periods as fundamentals and add the agentic layer at shares of 2\%, 5\% and 10\% (\cref{tab:hist}, \cref{fig:hist}). The layer deepens fast crashes most. At a 10\% share it adds \EnineCovidTen{}~bps to the 2020 drawdown, with a largest agent-induced decline of \EnineCovidMaxTen{}~bps; in 1998 it adds \EnineLTCMTen{}~bps and in 2011 \EnineEuroTen{}~bps. Slow bear markets behave differently. In 2008 and 2022 the layer produces large agent-induced declines along the way (\EnineGFCMaxTen{} and \EnineBearMaxTen{}~bps), but the final trough is not deeper (extra drawdown \EnineGFCTenSigned{} and \EnineBearTen{}~bps): agents de-risk early, front-loading the decline, and re-risk as volatility subsides before the fundamental trough. The policy-relevant risk is therefore acute: it is concentrated in fast, volatility-driven sell-offs of the kind seen in 1987, 2018 and 2020. M7 lowers the largest agent-induced decline by \EnineMaxRedPct\% on average across episodes and cuts the extra 2020 drawdown from \EnineCovidTen{} to \EnineCovidTenMSeven{}~bps.

\begin{table}[t]\centering\small
\caption{Historical counterfactuals. Actual daily constituent returns are replayed as fundamentals and the agentic layer is added at share $\alpha$. Fundamental drawdown: largest peak-to-trough decline of the equal-weighted index without agents (\%). Extra drawdown: change in that drawdown caused by the agentic layer (bps; negative values mean a shallower trough). Largest induced decline: maximum agent-induced price deviation during the episode (bps). Means over three seeds with 95\% intervals.}
\label{tab:hist}
\setlength{\tabcolsep}{4pt}\begin{tabular}{lrrrrrr}\toprule
 & & \multicolumn{3}{c}{Extra drawdown (bps)} & \multicolumn{2}{c}{Largest induced decline (bps)} \\ \cmidrule(lr){3-5}\cmidrule(lr){6-7}
Episode & Fund.\ DD (\%) & $\alpha=5\%$ & $\alpha=10\%$ & $10\%$+M7 & $\alpha=10\%$ & $10\%$+M7 \\ \midrule
1998 LTCM & 25.3 & $167 \pm 32$ & $253 \pm 61$ & $67 \pm 75$ & $561 \pm 85$ & $178 \pm 77$ \\
2000–01 dot-com & 26.3 & $79 \pm 35$ & $109 \pm 75$ & $-38 \pm 59$ & $351 \pm 25$ & $126 \pm 12$ \\
2008 GFC & 67.8 & $-98 \pm 24$ & $-148 \pm 26$ & $-111 \pm 26$ & $604 \pm 172$ & $319 \pm 188$ \\
2011 euro crisis & 18.8 & $140 \pm 33$ & $178 \pm 37$ & $31 \pm 77$ & $343 \pm 82$ & $70 \pm 71$ \\
2018 volatility shocks & 23.2 & $50 \pm 114$ & $88 \pm 149$ & $6 \pm 48$ & $312 \pm 82$ & $101 \pm 65$ \\
2020 COVID & 39.7 & $610 \pm 68$ & $858 \pm 112$ & $252 \pm 263$ & $939 \pm 102$ & $400 \pm 267$ \\
2022 bear market & 17.8 & $-38 \pm 42$ & $-15 \pm 100$ & $-15 \pm 18$ & $213 \pm 23$ & $65 \pm 12$ \\
\bottomrule\end{tabular}\end{table}

\section{A calibrated fragility index}\label{sec:index}

A supervisor cannot run our model on the real agentic layer: views, overlay parameters and client portfolios are private, and the agentic share of wealth is not known precisely. What a supervisor can observe, or could observe through the protocol of \cref{sec:privacy}, is the aggregate agentic flow $q_t$, public returns and an estimate of price impact. The amplification law \eqref{eq:law} combines exactly these quantities. We define the \emph{fragility index}
\begin{equation}\label{eq:fi}
\mathrm{FI} \;=\; 10^4\,\frac{\hat Y\,\hat\sigma\,\big[\mathrm{sd}(q)\big]^{\delta}}{1-\hat\kappa},\qquad
\hat\kappa=\hat\lambda\,\hat b,
\end{equation}
where $\mathrm{sd}(q)$ is the standard deviation of daily agentic index flow in units of daily volume, $\hat\sigma$ the daily volatility of public index returns, $\hat Y$ the impact prefactor, which brokers and supervisors routinely estimate from execution data \citep{toth2011,bouchaud2018}, and $\delta=1/2$. The loop gain $\hat\kappa$ is the product of two regressions on aggregates: $\hat\lambda$, the slope of index returns on same-day agentic flow (price impact), and $\hat b$, the slope of agentic flow on the previous day's index return (the layer's reaction to prices). Selling after declines ($\hat b<0$) combined with a negative price response to selling gives a positive gain. Every input is a sum or a public price, so FI can be computed without client-level data. The numerator measures how large synchronized trades are; the denominator how strongly they feed on themselves. The model enters only through the functional form; the index itself is a statistic of observable data.

\paragraph{Design.} To test whether FI is informative beyond the economies from which it was derived, we draw \EtwoN{} economies whose parameters vary at random over wide ranges: agentic share, number of providers and their shares, cross-model correlation, personalisation, prevalence and parameters of overlays, impact prefactor and decay, market turnover, extrapolation, attention, execution horizon, overlay heterogeneity and conviction (\cref{app:sim}). Each economy runs for five years. All predictors are computed from the first half of the economy and the target is $\SLaR_{99}$ in the second half, so the exercise is a forecast rather than a fit: a supervisor measures the layer today and must anticipate tail liquidation tomorrow. For each candidate measure $x$ we fit $\log\SLaR_{99}=a+b\log x$ on \EtwoTrain{} training economies and evaluate on \EtwoTest{} held-out economies; we also report five-fold cross-validation over all economies. The competing measures are those a supervisor might otherwise monitor: the agentic share $\alpha$; the flow volatility $\mathrm{sd}(q)$ alone; FI without the loop gain; the mean pairwise correlation of agent flows $\bar\rho_f$, itself privately measurable by Proposition~\ref{prop:private}; the model-level view correlation $(1-h)C$; and the provider Herfindahl index $H$.

\begin{table}[t]\centering\small
\caption{Out-of-sample performance of candidate fragility measures across \EtwoN{} randomly parameterised economies. Each predictor is computed from the first half of an economy (2.5 years) and used to forecast $\SLaR_{99}$ in the second half through a log--log regression fitted on \EtwoTrain{} training economies and evaluated on \EtwoTest{} held-out economies. CV $R^2$: five-fold cross-validation over all economies. MAE: mean absolute error of the implied forecast in bps.}
\label{tab:index}
\begin{tabular}{lrrrr}\toprule
Predictor & Hold-out $R^2$ & CV $R^2$ & Spearman $\rho$ & MAE (bps) \\ \midrule
fragility index FI & 0.56 & 0.55 & 0.80 & 39 \\
FI without loop gain & 0.56 & 0.55 & 0.79 & 39 \\
agentic flow volatility sd(q) & 0.35 & 0.35 & 0.63 & 49 \\
agentic share $\alpha$ & 0.23 & 0.24 & 0.50 & 51 \\
flow correlation $\bar\rho_f$ (private) & 0.01 & 0.04 & -0.03 & 60 \\
view correlation $(1-h)C$ & 0.12 & 0.05 & -0.38 & 55 \\
provider HHI $H$ & -0.01 & 0.00 & -0.02 & 59 \\
\bottomrule\end{tabular}\end{table}

\begin{figure}[t]
\centering
\includegraphics[width=\textwidth]{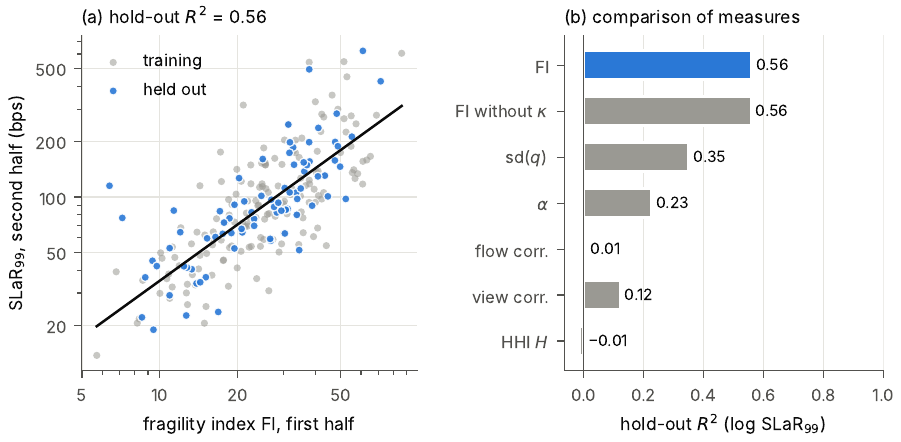}
\caption{The fragility index forecasts tail liquidation out of sample. (a) FI computed from the first half of each economy against $\SLaR_{99}$ in the second half, log scales; the line is the log--log fit on training economies. (b) Hold-out $R^2$ of each candidate measure.}
\label{fig:index}
\end{figure}

\paragraph{Results.} \EtwoIndexText

\paragraph{From index to basis points.} Because the forecast is a log--log regression, FI maps directly into a tail-loss forecast. The fitted elasticity of $\SLaR_{99}$ with respect to FI is \EtwoSlopeFI{}, and the mean absolute error of the implied forecast on held-out economies is \EtwoMAEFI{}~bps, against a median target of \EtwoTargetMed{}~bps (5th--95th percentiles \EtwoTargetLo--\EtwoTargetHi{}~bps). The index is therefore not only a ranking device but a calibrated quantity that a supervisor can compare with a tolerance---for example, the largest agent-induced five-day decline the market should absorb before its circuit breakers bind.

\paragraph{What the index does not do.} FI is calibrated to the model's impact law and to the range of economies in the ensemble. Its inputs are measurable, but its mapping into basis points should be re-estimated as aggregate agentic-flow data become available, and $\hat Y$ is itself uncertain. FI measures market-wide liquidation; the crowding in individual securities documented in \cref{sec:twokinds} requires asset-level and cross-asset flow statistics, which are sums of the same kind and can be computed by the same protocol.

\section{Mitigation: what decorrelates the layer, and at what cost}\label{sec:mitig}

\begin{figure}[t]
\centering
\includegraphics[width=\textwidth]{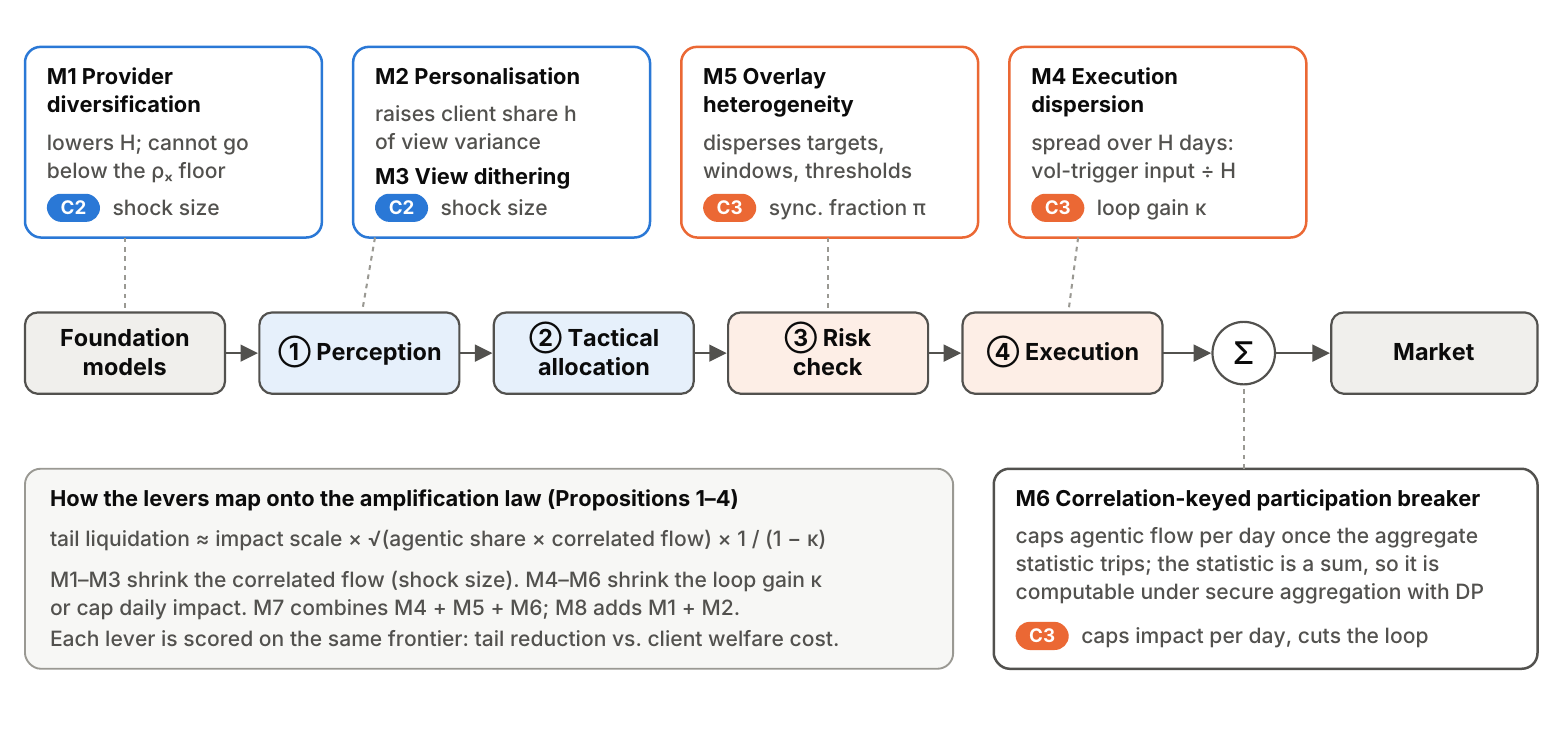}
\caption{Where each lever acts. View-level levers (M1--M3) shrink correlated demand; automation-level levers (M4--M6) shrink the loop gain or cap daily impact. M7 combines M4, M5 and M6; M8 adds M1 and M2.}
\label{fig:mitigmap}
\end{figure}

\begin{figure}[t]
\centering
\includegraphics[width=\textwidth]{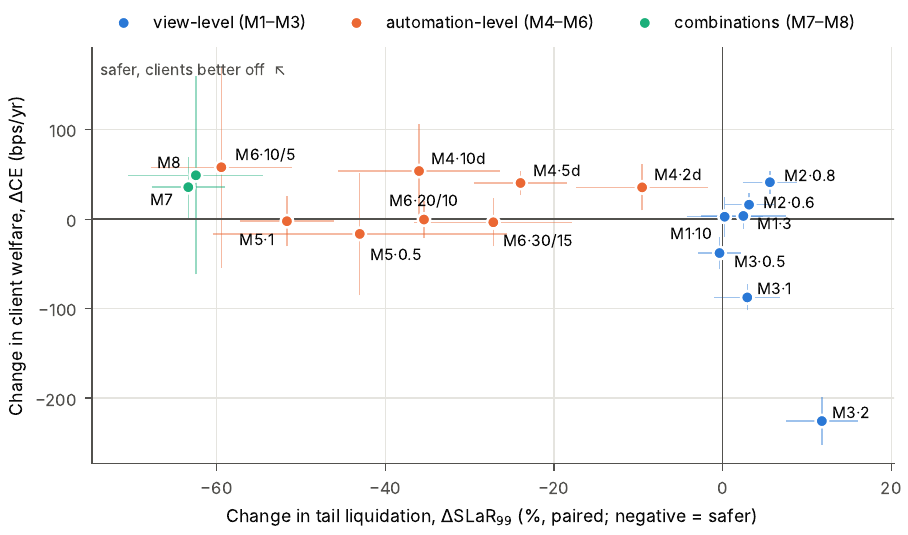}
\caption{Mitigation frontier. Paired change in $\SLaR_{99}$ (horizontal) against paired change in client certainty equivalent (vertical), with 95\% intervals; six paired seeds per lever and \EfiveSeeds{} for the baseline comparisons of M4 (five days), M5 (1.0), M6 (20/10) and M7 (\cref{tab:mitig}). Labels give the lever and its strength (e.g.\ M4$\cdot$5d: execution over five days; M6$\cdot$20/10: breaker trips at 20\% of volume and caps at 10\%).}
\label{fig:frontier}
\end{figure}

\Cref{fig:mitigmap} shows where each lever acts. M1--M3 act on views (the monoculture channel C2): provider diversification, more personalisation, and independent noise added to each agent's view (``input perturbation''). M4--M6 act on automation (C3): spreading execution over several days, dispersing overlay parameters (volatility targets, risk-model memory and drawdown thresholds), and a participation breaker that caps agentic flow per day once aggregate flow trips a threshold. M7 combines M4, M5 and M6; M8 adds M1 and M2 to M7. Every lever is evaluated against the same baseline on exactly paired simulations, so differences reflect the lever and not sampling noise. \Cref{tab:mitig} reports the results and \cref{fig:frontier} plots them on the frontier of tail reduction against client welfare.

\begin{table}[t]\centering\small
\caption{Mitigation frontier. Each lever is compared with the agentic baseline ($\alpha=5\%$, consumer concentration, $\rho_x=0.48$) on exactly paired simulations (same fundamentals, clients and views; five years each; the number of paired seeds is given in the last column). $\Delta\SLaR_{99}$ is the mean paired percentage change with a 95\% confidence interval; $\Delta$CE is the paired change in wealth-weighted certainty equivalent (bps per year). High-conviction rows (tactical tracking error 3\%) are compared with their own baseline.}
\label{tab:mitig}
\setlength{\tabcolsep}{4pt}\begin{tabular}{lrrrrr}\toprule
Lever & $\Delta\SLaR_{99}$ (\%) & $\Delta$CE (bps/yr) & Turnover & Breaker (\% days) & Seeds \\ \midrule
M1 provider diversification (10 equal) & $+0 \pm 5$ & $+3 \pm 22$ & 1.10 & 0.0 & 6 \\
M1 provider diversification (3 equal) & $+2 \pm 5$ & $+4 \pm 15$ & 1.09 & 0.0 & 6 \\
M2 personalisation, $h=0.6$ & $+3 \pm 3$ & $+16 \pm 14$ & 1.08 & 0.0 & 6 \\
M2 personalisation, $h=0.8$ & $+6 \pm 3$ & $+41 \pm 13$ & 1.09 & 0.0 & 6 \\
M3 view dithering, 0.5$\sigma$ & $-0 \pm 3$ & $-38 \pm 18$ & 3.45 & 0.0 & 6 \\
M3 view dithering, 1$\sigma$ & $+3 \pm 4$ & $-87 \pm 15$ & 6.49 & 0.0 & 6 \\
M3 view dithering, 2$\sigma$ & $+12 \pm 4$ & $-226 \pm 26$ & 11.33 & 0.0 & 6 \\
M4 execution over 2 days & $-10 \pm 8$ & $+36 \pm 26$ & 0.78 & 0.0 & 6 \\
M4 execution over 5 days & $-24 \pm 6$ & $+40 \pm 13$ & 0.60 & 0.0 & 18 \\
M4 execution over 10 days & $-36 \pm 10$ & $+54 \pm 53$ & 0.55 & 0.0 & 6 \\
M5 overlay heterogeneity, 0.5 & $-43 \pm 17$ & $-16 \pm 69$ & 1.02 & 0.0 & 6 \\
M5 overlay heterogeneity, 1.0 & $-52 \pm 6$ & $-2 \pm 28$ & 0.96 & 0.0 & 18 \\
M6 breaker, trip 10\%/cap 5\% of volume & $-59 \pm 8$ & $+58 \pm 112$ & 0.81 & 54.4 & 6 \\
M6 breaker, trip 20\%/cap 10\% & $-35 \pm 4$ & $-0 \pm 21$ & 0.95 & 14.6 & 18 \\
M6 breaker, trip 30\%/cap 15\% & $-27 \pm 9$ & $-3 \pm 27$ & 1.04 & 10.2 & 6 \\
M7 = M4 (5 days) + M5 (1.0) + M6 (20/10) & $-63 \pm 4$ & $+36 \pm 34$ & 0.57 & 0.7 & 18 \\
M8 = M7 + M1 (10) + M2 ($h=0.6$) & $-62 \pm 8$ & $+49 \pm 111$ & 0.62 & 0.8 & 6 \\
\midrule
High conviction: M1 (10 equal) & $+10 \pm 10$ & $+44 \pm 53$ & 2.96 & 0.0 & 6 \\
High conviction: M2 ($h=0.8$) & $+17 \pm 7$ & $+145 \pm 26$ & 2.97 & 0.0 & 6 \\
High conviction: M7 & $-61 \pm 12$ & $+117 \pm 167$ & 1.78 & 14.5 & 6 \\
\bottomrule\end{tabular}\end{table}

\paragraph{Automation-level levers work.} Spreading execution over five or ten days changes $\SLaR_{99}$ by \MFourFivedS\% and \MFourTendS\% and client welfare by \MFourFivedCE{} and \MFourTendCE{}~bps a year, because slower execution also lowers the impact costs that clients pay. Dispersing overlay parameters changes $\SLaR_{99}$ by \MFiveOnedS\% and welfare by \MFiveOnedCE{}~bps. The participation breaker changes $\SLaR_{99}$ by between \MSixCdS\% and \MSixAdS\% depending on how tight it is; the tightest version binds on \MSixABrk\% of days, the loosest on \MSixCBrk\%. The combination M7 changes $\SLaR_{99}$ by \MSevendS\% ($\pm$\MSevendSci) and client welfare by \MSevendCE{}~bps, and its breaker binds on only \MSevenBrk\% of days: once execution and overlays are dispersed, the breaker rarely needs to act.

\paragraph{View-level levers do not reduce index tails.} Diversifying providers changes $\SLaR_{99}$ by \MOneTendS\% ($\pm$\MOneTendSci) with ten equal providers and \MOneThreedS\% with three; neither is distinguishable from zero. This is what Proposition~\ref{prop:floor}(iv) and the risk floor of \citet{ross2026} lead one to expect when cross-model consensus is substantial, but the reason here is more basic, as we show next. More personalisation \emph{raises} $\SLaR_{99}$ slightly (\MTwoSixdS\% at $h=0.6$ and \MTwoEightdS\% at $h=0.8$) while improving welfare (\MTwoSixdCE{} and \MTwoEightdCE{}~bps). Injecting independent noise into views backfires: it multiplies turnover (to \MThreeTwoTurn{} at twice the view standard deviation), changes welfare by as much as \MThreeTwodCE{}~bps a year and $\SLaR_{99}$ by \MThreeTwodS\%. Adding diversification and personalisation to M7 (M8) leaves the tail reduction unchanged (\MEightdS\%) but adds welfare (\MEightdCE{}~bps). Among agents with high tactical conviction the pattern is the same: diversification changes $\SLaR_{99}$ by \HCMOnedS\%, personalisation by \HCMTwodS\%, and M7 by \HCMSevendS\%.

\subsection{Two kinds of fragility}\label{sec:twokinds}
\begin{figure}[t]
\centering
\includegraphics[width=\textwidth]{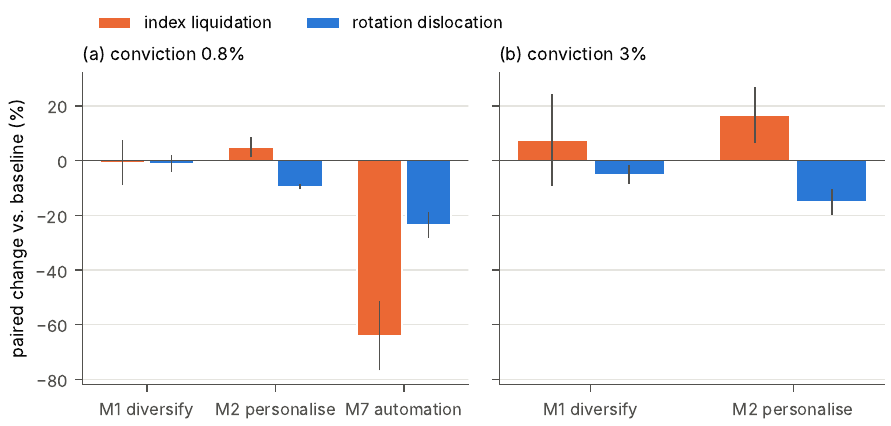}
\caption{Two kinds of fragility. Paired changes in index liquidation ($\SLaR_{99}$) and rotation dislocation (99th percentile of absolute five-day cross-sectional moves) relative to the baseline of the same tactical conviction; four seeds, 95\% intervals.}
\label{fig:twokinds}
\end{figure}

Why do view-level levers fail? Tactical views mostly move money \emph{between} assets: an agent that likes one stock buys it and sells others, so correlated views create crowded rotations but little net selling of the market as a whole. Index-level liquidation requires the whole layer to cut risky exposure at once, which is what overlays, drawdown controls and attention-driven execution do. We separate the two by recovering each asset's agent-induced move and splitting it into its index component and a rotation component (\cref{tab:twokinds}, \cref{fig:twokinds}).

\begin{table}[t]\centering\small
\caption{Two kinds of fragility. Index liquidation is $\SLaR_{99}$; single-asset dislocation is the 99th percentile of absolute five-day agent-induced moves across assets; rotation dislocation removes the index component from each asset's move. Levels in bps for the two baselines; other rows give paired percentage changes against the baseline of the same conviction (four seeds, 95\% intervals).}
\label{tab:twokinds}
\begin{tabular}{lrrr}\toprule
 & Index liquidation & Single-asset dislocation & Rotation dislocation \\ \midrule
Baseline, conviction 0.8\% (bps) & 195 & 390 & 327 \\
\quad M1 provider diversification (\%) & $-1 \pm 8$ & $+1 \pm 4$ & $-1 \pm 3$ \\
\quad M2 personalisation $h=0.8$ (\%) & $+5 \pm 4$ & $-4 \pm 6$ & $-10 \pm 1$ \\
\quad M7 automation combination (\%) & $-64 \pm 13$ & $-33 \pm 3$ & $-24 \pm 5$ \\
Baseline, conviction 3\% (bps) & 186 & 485 & 445 \\
\quad M1 provider diversification (\%) & $+8 \pm 17$ & $-3 \pm 5$ & $-5 \pm 3$ \\
\quad M2 personalisation $h=0.8$ (\%) & $+17 \pm 10$ & $-9 \pm 7$ & $-15 \pm 5$ \\
\bottomrule\end{tabular}\end{table}

The decomposition shows two distinct fragilities. Raising tactical conviction from 0.8\% to 3\% tracking error increases rotation dislocations from \ALBaseRot{} to \ALHCRot{}~bps and single-asset dislocations from \ALBaseAsset{} to \ALHCAsset{}~bps, but leaves index liquidation essentially unchanged (\ALBaseIdx{} versus \ALHCIdx{}~bps). View monoculture is therefore a driver of \emph{cross-sectional crowding}: many clients piling into, or out of, the same names. View-level levers address that fragility---personalisation changes rotation dislocations by \ALHCMTwoRot\% among high-conviction agents---but they do so while slightly \emph{raising} index liquidation (\ALHCMTwoIdx\%). The reason is the concavity of impact: rotation flows leave assets with offsetting imbalances that partly absorb the common de-risking flow, so removing them leaves the common flow to hit thinner liquidity. Automation-level levers address both fragilities: M7 changes index liquidation by \ALMSevenIdx\%, single-asset dislocations by \ALMSevenAsset\% and rotation dislocations by \ALMSevenRot\%.

The policy implication is direct. If the concern is a market-wide liquidation, decorrelating the models is the wrong lever; decorrelating the automation is the right one. If the concern is crowding in individual securities, model diversity and personalisation help.

\subsection{Which lever for which overlay?}\label{sec:prop4test}
\begin{figure}[t]
\centering
\includegraphics[width=0.62\textwidth]{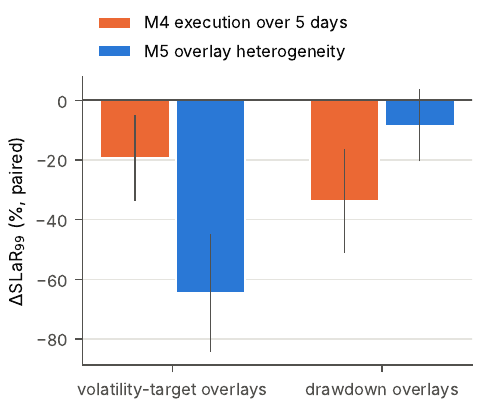}
\caption{Which lever for which overlay. Paired change in $\SLaR_{99}$ from execution dispersion (M4, five days) and overlay heterogeneity (M5) in economies where all agents use volatility targets (left) or drawdown controls (right); $\alpha=10\%$, six seeds, 95\% intervals.}
\label{fig:prop4}
\end{figure}

Proposition~\ref{prop:dither}, read naively, assigns execution dispersion to volatility-triggered de-risking and threshold dispersion to loss-triggered de-risking. We test this directly by running economies in which every agent uses a volatility target and none a drawdown control, and the reverse, each at $\alpha=10\%$ (\cref{fig:prop4}).

The full model reverses the naive assignment. Against volatility-target overlays, heterogeneity changes $\SLaR_{99}$ by \EtenVtMFive\% ($\pm$\EtenVtMFiveCI) and execution dispersion by \EtenVtMFour\% ($\pm$\EtenVtMFourCI). Against drawdown overlays, execution dispersion changes it by \EtenDdMFour\% ($\pm$\EtenDdMFourCI) and heterogeneity by only \EtenDdMFive\% ($\pm$\EtenDdMFiveCI), which is not distinguishable from zero. Two features of realistic markets explain the reversal, both outside the assumptions of Proposition~\ref{prop:dither}. Volatility targets are triggered mostly by common \emph{fundamental} volatility, observed by every agent through the same risk model (channel C1); slower execution cannot change that signal, but dispersing how agents respond to it can. Drawdown controls produce large, discrete exposure cuts; with decaying impact and volatility-dependent liquidity, spreading them lowers the cumulative loss they cause and hence the further triggers they set off, while their thresholds are already dispersed by clients' different strategic allocations. The lesson for platform design is specific: disperse the parameters of volatility-based rules, and slow down the execution of loss-based rules.

\subsection{Robustness}\label{sec:robust}
\Cref{tab:robust} in the appendix and \cref{fig:robust} re-run the baseline and M7 under \EeightNcfg{} alternative assumptions: linear impact, half and one-and-a-half times the impact prefactor, slower and faster impact decay, low and high market turnover, additional liquidity withdrawal, no or triple extrapolation, universal overlays, smaller and larger agent populations, and short bootstrap blocks. M7 lowers $\SLaR_{99}$ in every case, by between \EeightMinRed\% and \EeightMaxRed\%. Its welfare effect is less robust: in these three-seed runs the paired change in the certainty equivalent ranges from \EeightCEMin{} to \EeightCEMax{}~bps a year and \EeightCESig{} of the estimates is distinguishable from zero. The robust conclusion is therefore that M7 removes about two-thirds of tail liquidation at no measurable cost to clients, rather than that it makes them better off.

\begin{figure}[t]
\centering
\includegraphics{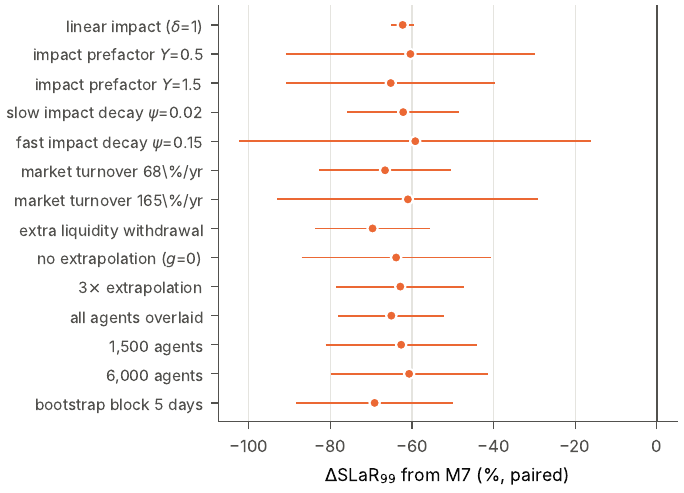}
\caption{Robustness of the combined mitigation M7: paired change in $\SLaR_{99}$ under alternative assumptions (three seeds, 95\% intervals).}
\label{fig:robust}
\end{figure}

\section{Measuring and acting on correlation without pooling client data}\label{sec:privacy}

\begin{figure}[t]
\centering
\includegraphics[width=\textwidth]{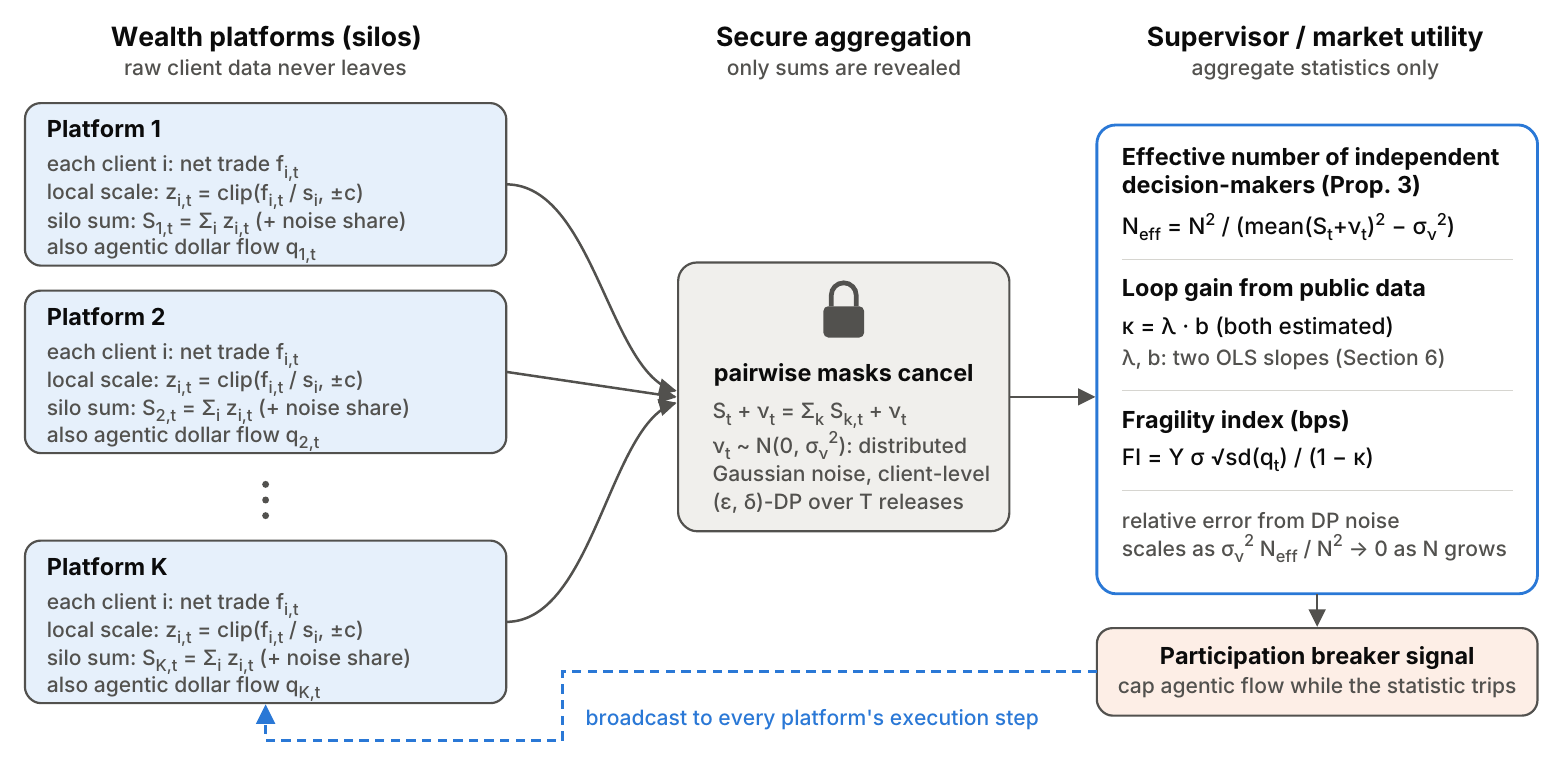}
\caption{Privacy-preserving measurement. Platforms compute standardised flow sums locally; secure aggregation reveals only the noisy market-wide sum; the supervisor computes $\Neff$, the loop gain and the fragility index, and broadcasts the breaker signal. Client-level $(\varepsilon,\delta)$-differential privacy holds over all releases.}
\label{fig:private}
\end{figure}

The fragility index and the participation breaker both need cross-platform information that no single firm holds and that no firm can share: the trades of its clients. Proposition~\ref{prop:private} shows that the statistics required are sums. \Cref{fig:private} sketches a protocol. Each platform standardises each client's net trade locally, clips it, and adds the results into a platform sum; platforms combine their sums by secure aggregation, so that the supervisor or a market utility learns only the market-wide total plus calibrated Gaussian noise \citep{bonawitz2017,dwork2014}. The same channel carries the aggregate dollar flow $q_t$ that enters the fragility index and triggers the breaker. Public returns supply the remaining inputs.

\begin{figure}[t]
\centering
\includegraphics[width=\textwidth]{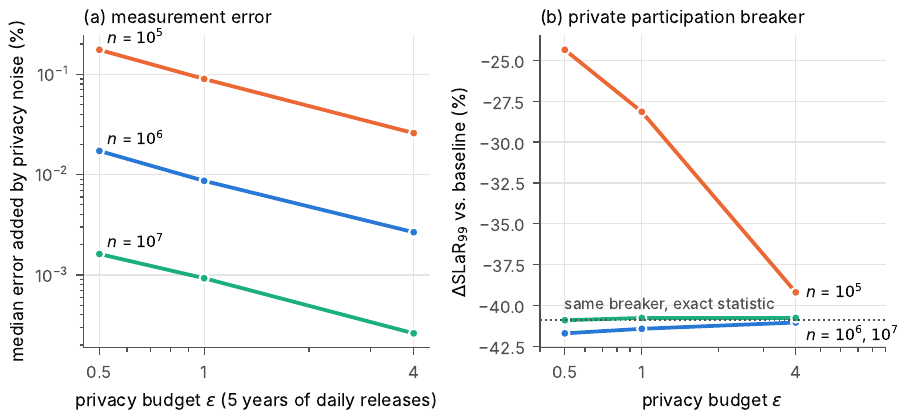}
\caption{Privacy-preserving measurement and control. (a) Median relative error of the private $\Neff$ estimate against the privacy budget for different client populations. (b) Paired change in $\SLaR_{99}$ from a participation breaker driven by the private statistic; the dotted line is the same breaker on the exact statistic.}
\label{fig:privacy}
\end{figure}

\paragraph{Measurement accuracy.} We take the standardised flow sums produced by the baseline economies, treat each simulated agent as representing $n/N$ real clients, add the Gaussian noise required for client-level privacy over all \num{1260} daily releases, and re-estimate $\Neff$ (\cref{fig:privacy}a). With $10^5$ clients and the strictest budget ($\varepsilon=0.5$), the median error of the private estimate is \EsevenNeffErrAHalf\%; with $10^7$ clients it falls to \EsevenNeffErrCHalf\%, as Proposition~\ref{prop:private} predicts. At the scale of today's AI-using retail population, systemic correlation can be measured with negligible privacy cost.

\paragraph{A private breaker.} \Cref{tab:privacy} and \cref{fig:privacy}b run the participation breaker on the noisy aggregate. On the same seeds, the breaker driven by the exact statistic changes $\SLaR_{99}$ by \EsevenTrueBrk\%. With $10^7$ clients the private breaker matches it at every budget (\EsevenCHalfdS\% at $\varepsilon=0.5$) with \EsevenCHalfFp{} percentage points of extra trips. With $10^5$ clients and $\varepsilon=0.5$, noise trips the breaker on many additional days (\EsevenAHalfFp{} percentage points), which still lowers tail risk (\EsevenAHalfdS\%) but by throttling ordinary trading; a supervisor operating at that scale would choose a larger budget or aggregate over a longer window.

\begin{table}[t]\centering\small
\caption{Participation breaker driven by a differentially private aggregate (trip at 20\% of volume, cap 10\%). Privacy is client-level $(\varepsilon,10^{-5})$ over all 1{,}260 daily releases; $n$ is the number of clients behind the aggregate. $\Delta\SLaR_{99}$: paired change against the unmitigated baseline (the same breaker on the exact statistic achieves \EsevenTrueBrk\% on the same seeds). Extra trips: additional breaker days caused by noise, in percentage points.}
\label{tab:privacy}
\begin{tabular}{rrrr}\toprule
Clients $n$ & $\varepsilon$ & $\Delta\SLaR_{99}$ (\%) & Extra trips (pp) \\ \midrule
$10^{5}$ & 0.5 & $-24 \pm 15$ & $23.4$ \\
$10^{5}$ & 1 & $-28 \pm 13$ & $18.6$ \\
$10^{5}$ & 4 & $-39 \pm 11$ & $2.6$ \\
$10^{6}$ & 0.5 & $-42 \pm 8$ & $-1.0$ \\
$10^{6}$ & 1 & $-41 \pm 8$ & $-0.6$ \\
$10^{6}$ & 4 & $-41 \pm 7$ & $0.0$ \\
$10^{7}$ & 0.5 & $-41 \pm 7$ & $0.0$ \\
$10^{7}$ & 1 & $-41 \pm 7$ & $0.0$ \\
$10^{7}$ & 4 & $-41 \pm 7$ & $0.0$ \\
\bottomrule\end{tabular}\end{table}

\section{Policy implications}\label{sec:policy}

\paragraph{Measure the layer, not the firm.} The relevant unit of risk is the agentic layer as a whole: an individual platform can be well diversified and well controlled while the layer behaves like a few investors. Supervisors already collect transaction data; what is new is the need for a cross-platform statistic of synchronized flow. Because the statistics are sums, they can be produced by a market utility under secure aggregation with formal privacy guarantees (\cref{sec:privacy}). The fragility index of \cref{sec:index} is one such statistic; it requires the aggregate flow, public returns and an impact estimate.

\paragraph{Target automation, not only model concentration.} Model concentration and cross-model consensus are real (\cref{sec:theory}) and they matter for crowding in individual securities and for shared-misinformation shocks (\cref{sec:twokinds,sec:misinfo}). But the market-wide liquidations we document are driven by synchronized automation: risk overlays that react to the same volatility through the same risk models, drawdown cuts executed at once, and activity that spikes when markets move. Provider diversification does not reduce them. Requirements that are cheap for clients do: dispersion of overlay parameters, staged execution of large exposure cuts, and participation limits that bind only in stress. In our experiments the combination lowers tail liquidation by about two-thirds at no measurable cost to client welfare.

\paragraph{Learn from portfolio insurance.} After 1987, the response focused on circuit breakers and on understanding mechanical selling \citep{brady1988,greenwald1991}. The agentic wealth layer reproduces the 1987 configuration---a common formula, executed mechanically, by a large share of the market---but at retail scale and at the speed of a chat. Two lessons carry over directly. Market-wide circuit breakers remain necessary but act late; and the magnitude of formula-driven selling depends on how many participants run the formula at once, which is a design choice of the platforms \citep{brady1988,gennotte1990}.

\paragraph{Mind the conflict between suitability and stability.} Overlays exist to protect clients: a volatility target keeps risk within a client's tolerance and a drawdown control caps losses. Both are sensible for each client and fragile in aggregate. Dispersion requirements preserve the protection---every client still has a risk limit---while breaking its synchronization. Supervisors and platforms should treat the parameters of client-protection rules as a macroprudential variable.

\section{Limitations}\label{sec:limits}

The model is a calibrated simulation, not an estimate from data on agentic flows, which do not yet exist at scale. Several simplifications matter. Assets have equal capitalisation and the universe is 20 large-cap U.S. equities, so the model is silent on bonds, credit, foreign exchange and less liquid assets, where impact is larger. Non-agentic investors are represented only through the latent-liquidity impact function; strategic liquidity providers who anticipate agentic flows, or arbitrageurs who trade against them, could either dampen or front-run the liquidations we describe. Agents follow rules rather than being language models; we calibrate the correlation of their views to measured correlations among real models \citep{ross2026,chen2024}, but language models may also correlate through channels we do not model, such as shared retrieval sources. The adoption share $\alpha$ is a scenario variable. The welfare measure is a certainty equivalent over five-year paths and ignores benefits of agentic advice that are not captured by returns, such as access and engagement \citep{dacunto2019}. Finally, the privacy analysis covers client-level differential privacy of the released statistics and assumes the secure-aggregation protocol is implemented correctly; it does not address collusion between platforms and the aggregator beyond the standard threat model.

\section{Conclusion}\label{sec:concl}

\ConclusionText

\section*{Data and code availability}
All code, the processed real-return data and every raw simulation result are released with this paper. Real returns are from the open-source \texttt{skfolio} datasets \citep{skfolio}; all other data are generated by the documented simulator from fixed seeds. A single script regenerates every table, figure and number.

\section*{Declaration of competing interests}
The author declares no competing interests and received no funding for this work.

\section*{Declaration of generative AI and AI-assisted technologies in the research and writing process}
\AIDeclarationText

\bibliographystyle{plainnat}
\bibliography{references}

\appendix
\section{Proofs}\label{app:proofs}

\paragraph{Proof of Proposition~\ref{prop:floor}.} Write $v_i = s_v[\sqrt{1-h}\,b_{m(i)}+\sqrt h\,u_i]$ with $b_m=\sqrt{\rho_x}\beta+\sqrt{1-\rho_x}\eta_m$, where $\beta$, $\eta_m$ and $u_i$ are mutually independent with mean zero and unit variance. Then $\Var(v_i)=s_v^2$. For $i\neq j$ served by the same provider, $\Cov(v_i,v_j)=s_v^2(1-h)\Var(b_m)=s_v^2(1-h)$; for different providers, $\Cov(v_i,v_j)=s_v^2(1-h)\rho_x$. Among the $N(N-1)$ ordered pairs, a fraction $\tilde H=\sum_m n_m(n_m-1)/[N(N-1)]$ share a provider, so
$\bar\rho=(1-h)[\tilde H+\rho_x(1-\tilde H)]=(1-h)[\rho_x+(1-\rho_x)\tilde H]$, which is (i). For (ii),
\[
\bar x=\frac aN\sum_i v_i = a s_v\Big[\sqrt{1-h}\Big(\sqrt{\rho_x}\beta+\sqrt{1-\rho_x}\sum_m \tfrac{n_m}{N}\eta_m\Big)+\sqrt h\,\tfrac1N\sum_i u_i\Big],
\]
and independence gives $\Var(\bar x)=a^2s_v^2[(1-h)(\rho_x+(1-\rho_x)\sum_m(n_m/N)^2)+h/N]$. For (iii), $n_m/N\to s_m$ implies $\tilde H\to H$ and $\bar\rho\to(1-h)C>0$, so $\Neff=N/[1+(N-1)\bar\rho]\to 1/\bar\rho$. Part (iv) follows from $H_N\ge0$. \hfill$\square$

\paragraph{Proof of Proposition~\ref{prop:mult}.} Consider rounds $j=0,1,\dots$ shorter than the time over which impact decays. A move $\Delta p_j$ changes the value of a loss-linear overlay's risky holding $E=m(W-\Phi)$ (wealth $W$, floor $\Phi$) by $E\Delta p_j$; the desired holding becomes $E+mE\Delta p_j$ while the existing holding has drifted to $E(1+\Delta p_j)$, so the trade per unit of agentic wealth is $(m-1)E\Delta p_j=\varepsilon\Delta p_j$. Aggregating over the synchronized fraction $\pi$ of agentic wealth, a share $\alpha$ of the market, the net trade is $\pi\alpha\varepsilon\Delta p_j$ and its impact is $\Delta p_{j+1}=\lambda\pi\alpha\varepsilon\Delta p_j/D=\kappa\Delta p_j$. The cumulative move $\sum_{j\ge0}\kappa^j\Delta p_0$ equals $\Delta p_0/(1-\kappa)$ if $|\kappa|<1$ and diverges otherwise; $\kappa=1$ at $\alpha^\star=D/(\pi\varepsilon\lambda)$. The result is exact to first order in $\Delta p_0$; \cref{tab:theory} confirms it numerically. Under \eqref{eq:impact} the local impact slope is $\partial d/\partial F=\delta Y\hat\sigma|F|^{\delta-1}$, decreasing in $|F|$ for $\delta<1$ and increasing in $\hat\sigma$. \hfill$\square$

\paragraph{Proof of Proposition~\ref{prop:private}.} $\E[S_t^2]=\sum_i\Var(z_{i,t})+\sum_{i\neq j}\Cov(z_{i,t},z_{j,t})=N+N(N-1)\bar\rho$, hence $N/[1+(N-1)\bar\rho]=N^2/\E[S_t^2]$. Under stationarity and ergodicity $T^{-1}\sum_t\tilde S_t^2\to\E[S_t^2]+\sigma_\nu^2$ in probability, and the continuous-mapping theorem gives consistency. If $S_t\sim\mathcal N(0,v)$ with $v=\E[S_t^2]$, independent over $t$ and of $\nu_t$, then $\tilde S_t^2$ has mean $v+\sigma_\nu^2$ and variance $2(v+\sigma_\nu^2)^2$, so $\hat M=T^{-1}\sum_t\tilde S_t^2-\sigma_\nu^2$ has relative standard error $\sqrt{2/T}(1+\sigma_\nu^2/v)$; the delta method transfers this to $\hat N_{\mathrm{eff}}=N^2/\hat M$, and $\sigma_\nu^2/v=\sigma_\nu^2\Neff/N^2$. For privacy, adding or removing one client changes $S_t$ by at most $c$, so each release is a Gaussian mechanism with noise multiplier $z=\sigma_\nu/c$ and R\'enyi divergence $a/(2z^2)$ at order $a$ \citep{mironov2017}; R\'enyi divergences add under composition, and the stated conversion to $(\varepsilon,\delta)$ follows \citet{canonne2020}. \hfill$\square$

\paragraph{Proof of Proposition~\ref{prop:dither}.} Each of the $H$ periods carries a move $\lambda X/H$. To first order in $\vartheta$ the EWMA variance increment is $\vartheta\sum_{j=1}^H(\lambda X/H)^2=\vartheta\lambda^2X^2/H$, against $\vartheta\lambda^2X^2$ for immediate execution, which gives (i) because volatility-target exposure is a smooth decreasing function of the variance estimate. The cumulative move $\sum_j\lambda X/H=\lambda X$ does not depend on $H$, which gives (ii) because drawdown triggers depend on cumulative loss. With geometric decay the deviation after the last child order is $\sum_{j=0}^{H-1}(1-\psi)^j\lambda X/H=\lambda X[1-(1-\psi)^H]/(\psi H)$, its maximum, which gives (iii). \hfill$\square$

\section{Simulation details}\label{app:sim}

\paragraph{Daily algorithm.} For $t=1,\dots,T_{\mathrm{burn}}+T$: (1) update the AR(1) shocks $\beta_t,\eta_{m,t},u_{i,t}$ and form views \eqref{eq:view}, adding misinformation or dither noise when an experiment requires it; (2) compute tactical targets from the shared shrunk covariance; (3) draw conversations and refresh the tactical targets of conversing agents; (4) apply volatility-target and drawdown overlays to all agents; (5) form trades subject to the band and execution horizon; (6) aggregate flow into $q_{k,t}$, apply the participation breaker (if enabled) and the one-day-volume capacity cap; (7) update $F_{k,t}$ and $d_{k,t}$ by \eqref{eq:impact}, enforcing the 7\% market-wide breaker by bisection on the agentic flow; (8) settle trades and costs, update wealth, peaks, holdings, the EWMA covariance and the trailing return; (9) record aggregates, including the locally standardised flow sum $S_t$ used in \cref{sec:privacy}. Agents' own flow scales $s_i$ are exponentially weighted moving averages computed locally.

\paragraph{Ensemble design for the fragility index.} \EnsembleDesignText

\paragraph{Computation and reproducibility.} The model is written in Python with NumPy. A five-year economy with \num{3000} agents runs in about eight seconds on one CPU core. In total the paper reports \TotalRuns{} simulated economies. Code, the processed real-return data and every raw result are released with the paper; each table and figure is produced by scripts from those raw results, and every number in the text is generated from them.

\begin{table}[t]\centering\small
\caption{Robustness of the combined mitigation M7. Each row changes one assumption of the baseline; $\SLaR_{99}$ is the baseline value under that assumption and $\Delta$ columns give paired changes from adding M7 (three seeds, 95\% intervals). Extra liquidity withdrawal multiplies the impact of each day's flow by $\min\{\hat\sigma^f_t/\bar\sigma,2\}$, where $\hat\sigma^f_t$ is the EWMA volatility of fundamental returns, so that liquidity providers retreat in market stress \citep{nagel2012}. With a cap of 3 instead of 2, one of the three baseline economies collapses (agent-induced deviations grow without bound) while all three M7 economies remain stable.}
\label{tab:robust}
\begin{tabular}{lrrr}\toprule
Variation & $\SLaR_{99}$ (bps) & $\Delta\SLaR_{99}$ (\%) & $\Delta$CE (bps/yr) \\ \midrule
linear impact ($\delta$=1) & 237 & $-62 \pm 3$ & $-48 \pm 132$ \\
impact prefactor $Y$=0.5 & 93 & $-60 \pm 31$ & $-80 \pm 134$ \\
impact prefactor $Y$=1.5 & 309 & $-65 \pm 26$ & $+55 \pm 182$ \\
slow impact decay $\psi$=0.02 & 203 & $-62 \pm 14$ & $-71 \pm 117$ \\
fast impact decay $\psi$=0.15 & 157 & $-59 \pm 43$ & $-38 \pm 149$ \\
market turnover 68\%/yr & 235 & $-67 \pm 16$ & $-43 \pm 110$ \\
market turnover 165\%/yr & 143 & $-61 \pm 32$ & $-69 \pm 145$ \\
extra liquidity withdrawal & 247 & $-70 \pm 14$ & $+273 \pm 886$ \\
no extrapolation ($g$=0) & 200 & $-64 \pm 23$ & $-62 \pm 124$ \\
3$\times$ extrapolation & 179 & $-63 \pm 16$ & $-39 \pm 153$ \\
all agents overlaid & 283 & $-65 \pm 13$ & $-103 \pm 194$ \\
1,500 agents & 189 & $-63 \pm 19$ & $-57 \pm 135$ \\
6,000 agents & 193 & $-61 \pm 19$ & $-56 \pm 146$ \\
bootstrap block 5 days & 166 & $-69 \pm 19$ & $+37 \pm 63$ \\
\bottomrule\end{tabular}\end{table}

\end{document}